\documentclass[sigconf, nonacm]{acmart}

\newcommand{\eat}[1]{}

\usepackage{lipsum}
\usepackage{latexsym}
\usepackage{amsfonts}
\usepackage{amsmath}

\usepackage{algorithm}
\usepackage{algorithmicx}
\usepackage{algpseudocode}

\usepackage{xcolor}
\usepackage{colortbl}
\usepackage{epsfig}
\usepackage{xspace}
\usepackage{graphicx}
\usepackage{paralist}
\usepackage{enumerate}
\usepackage{enumitem}
\usepackage[color,matrix,arrow,all]{xy}
\usepackage{comment}
\usepackage{booktabs}
\usepackage{balance}
\usepackage{stmaryrd}
\usepackage{pifont}
\usepackage{hhline}
\usepackage{listings}
\usepackage{array}
\usepackage{float}
\usepackage[flushleft]{threeparttable}

\usepackage{mathrsfs}
\usepackage{makecell}
\usepackage{xparse}
\usepackage{wrapfig}

\usepackage{tabularx}
\usepackage{subcaption}

\usepackage{epsfig}
\usepackage{multirow}
\usepackage{url}

\usepackage{multirow}
\usepackage{natbib}
\usepackage{graphicx}

\usepackage[most]{tcolorbox}

\usepackage{listings}
\usepackage{framed}
\usepackage{xcolor}
\usepackage{color}
\usepackage{changepage}
\colorlet{shadecolor}{gray!20}
\usepackage{makecell}

\usepackage{fontawesome5} 
\usepackage{tabularray}

\usepackage{amsmath}
\DeclareMathOperator*{\argmax}{arg\,max}

\definecolor{shadecolor}{RGB}{220,220,220}

\definecolor{inputcolor}{RGB}{255,139,35}
\definecolor{outputcolor}{RGB}{120,212,252}
\definecolor{embedcolor}{RGB}{254,127,156}
\definecolor{maskcolor}{RGB}{122,128,255}
\definecolor{ecolor}{RGB}{58,149,54}

\definecolor{highcolor}{RGB}{255,153,153}
\definecolor{midcolor}{RGB}{255,204,204}
\definecolor{lowcolor}{RGB}{204,229,255}

\usepackage{tikz}
\usetikzlibrary{shapes,snakes}
\usetikzlibrary{calc}

\usepackage[export]{adjustbox}

\definecolor{green}{RGB}{0,128,0}

\definecolor{yellow}{RGB}{255,200,18}

\newcommand{\stab}{\vspace{1.2ex}\noindent}

\newcommand{\be}{\begin{enumerate}}
\newcommand{\ee}{\end{enumerate}}
\newcommand{\beqn}{\begin{eqnarray*}}
\newcommand{\eeqn}{\end{eqnarray*}}

\newcommand{\stitle}[1]{\stab\noindent{\bf #1}}
\newcommand{\ptitle}[1]{\vspace{1.0mm}\noindent{\bf #1}}

\newcommand{\ie}{\textit{i.e.,}\xspace}
\newcommand{\eg}{\textit{e.g.,}\xspace}

\makeatletter
    \newcommand\figcaption{\def\@captype{figure}\caption}
    \newcommand\tabcaption{\def\@captype{table}\caption}
\makeatother

\tikzstyle{mybox} = [draw=black, fill=black!5, thick,
   rectangle, rounded corners, inner sep=0pt, inner ysep=6pt]
\tikzstyle{fancytitle} =[fill=black, text=white]

\renewcommand{\[}{\begin{equation}}
\renewcommand{\]}{\end{equation}}

\definecolor{InsightBlue}{RGB}{60, 96, 170}
\definecolor{InsightBg}{RGB}{245, 248, 255}
\definecolor{InsightBorder}{RGB}{180, 198, 231}

\newtcolorbox{insightbox}{
  enhanced,
  colback=InsightBg,
  colframe=InsightBorder,
  boxrule=0.5pt,
  arc=2.5pt,
  left=2pt,
  right=2pt,
  top=1pt,
  bottom=1pt,
  borderline west={2pt}{0pt}{InsightBlue},
  before skip=4pt,
  after skip=6pt
}

\newcommand{\nlq}{{\rm NL}\xspace}
\newcommand{\sql}{{\rm SQL}\xspace}

\newcommand{\nlsql}{{Text-to-SQL}\xspace}

\newcommand{\sys}{\text{SQLConductor}\xspace}

\definecolor{sqlkeyword}{rgb}{0, 0, 0.8}   
\definecolor{sqlstring}{rgb}{0.8, 0.4, 0}    
\definecolor{sqlcomment}{rgb}{0, 0.5, 0}   

\lstdefinestyle{sqlstyle}{
  language=SQL,
  basicstyle=\ttfamily\scriptsize,
  keywordstyle=\color{sqlkeyword}\bfseries,
  stringstyle=\color{sqlstring},
  commentstyle=\color{sqlcomment}\itshape,
  breaklines=true,
  breakindent=0pt,
}

\algrenewcommand{\algorithmiccomment}[1]{\bgroup\color{gray}\textit{// #1}\egroup}
\algrenewcommand{\algorithmicrequire}{\textbf{Input:}}
\algrenewcommand{\algorithmicensure}{\textbf{Output:}}
\algnewcommand{\Func}[1]{\mathtt{#1}}

\definecolor{pzyyellow}{HTML}{FADA5E}

\setlist[itemize]{leftmargin=0.35cm, topsep=0pt, itemsep=0pt}
\setlist[enumerate]{leftmargin=0.35cm, topsep=0pt, itemsep=0pt}

\definecolor{groupgray}{RGB}{245,245,245}
\definecolor{oursgreen}{RGB}{226,239,218}

\begin{document}

\title{\sys: Search-to-Policy Learning for Step-wise Text-to-SQL Orchestration} 

\author{Yizhang Zhu}
\affiliation{%
  \institution{HKUST(GZ)}
  \city{Guangzhou}
  \state{China}
}
\email{yzhu305@connect.hkust-gz.edu.cn}

\author{Zhangyang Peng}
\affiliation{%
  \institution{HKUST(GZ)}
  \city{Guangzhou}
  \state{China}
}
\email{zpeng529@connect.hkust-gz.edu.cn}

\author{Boyan Li}
\affiliation{%
  \institution{HKUST(GZ)}
  \city{Guangzhou}
  \state{China}
}
\email{bli303@connect.hkust-gz.edu.cn}

\author{Yuyu Luo*}
\affiliation{%
  \institution{HKUST(GZ)}
  \city{Guangzhou}
  \state{China}
}
\email{yuyuluo@hkust-gz.edu.cn}

\begin{abstract}

\nlsql enables users to access relational databases through natural language questions, but real-world \nlsql remains challenging due to the need for coordinated reasoning over complex database environments.
Existing systems often rely on carefully designed multi-stage \nlsql pipelines, sometimes further enhanced with trained reasoning models specialized for individual stages.
However, such fixed pipelines rely on predefined stage orders, limiting their adaptivity to different query demands and intermediate evidence.
Recent orchestration-based methods provide a more flexible alternative by composing specialized modules for each query, but typical plan-then-execute orchestration still commits to a complete workflow before execution and cannot sufficiently adapt to intermediate artifacts and feedback.

In this paper, we propose \sys, a \textit{step-wise orchestration} learning framework for \nlsql. 
\sys formulates typical \nlsql subtasks as specialized actions in an action space for flexible workflow composition, and trains a policy model to decide the next action based on intermediate artifacts and feedback.
To learn such an adaptive policy, \sys introduces \textit{Search-to-Policy Learning}, which uses Monte Carlo Tree Search to explore candidate workflows and stability estimation to identify robust workflow supervision. 
The policy model is trained with Stability-weighted Supervised Fine-tuning to prioritize high-quality orchestration patterns, and further enhanced through Curriculum Reinforcement Learning.
In this way, broad offline workflow search is transformed into a deployable policy for efficient step-wise orchestration at inference time.
Extensive experiments on BIRD-Dev and out-of-distribution datasets show that \sys achieves superior execution accuracy and strong generalization, reaching 73.2\% EX on BIRD-Dev by training a compact orchestration policy to coordinate frozen larger action models, outperforming prior methods directly training comparable or larger \nlsql backbones. 
Further analyses show that the learned policy adapts workflow orchestration to diverse query demands.

\end{abstract}

\maketitle


\section{Introduction}

\nlsql aims to enable non-technical users to access relational databases through natural language (\nlq) questions~\cite{liu2025survey, hong2025next} and has become an important interface for data analytics and business intelligence in the era of large language models (LLMs)~\cite{dataagent_survey, statqa, nvbench2, deepeye_icde}.
Despite recent progress, real-world \nlsql remains challenging.
A practical \nlsql system needs to locate relevant tables and columns in large and heterogeneous schemas, align query conditions with database contents, and construct complex query logic~\cite{supersql, nl2sql_tutorial}.
Correct \sql generation therefore requires coordinating these reasoning steps into an effective solving procedure.

\begin{figure}[t!]
    \centering
    \includegraphics[width=0.98\columnwidth]{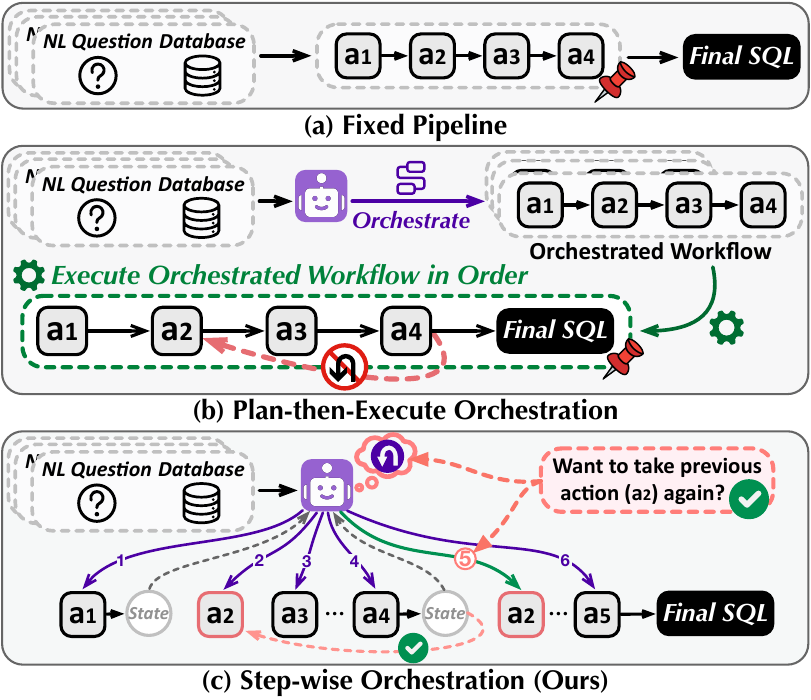}
    \vspace{-1.0em}
    \caption{Comparison of fixed workflow, plan-then-execute orchestration, and step-wise orchestration for \nlsql.}
    \label{fig:nlsql_compare}
    \vspace{-1.5em}
\end{figure}

\stitle{From Fixed Pipelines to Orchestration.}
To handle these difficulties, some \nlsql systems introduce meticulously designed pipelines to decompose \nlsql into more manageable subtasks, such as schema linking, value retrieval, \sql generation, revision, and selection~\cite{c3sql, macsql, rslsql, opensearchsql}.
Some methods further strengthen such pipelines by incorporating stage-specific reasoning models trained with large-scale supervision~\cite{opensql, cscsql, agentarscalesql}.
These designs have achieved strong performance by introducing structured reasoning procedures and task-specific adaptation.

However, as shown in Figure~\ref{fig:nlsql_compare}(a), this paradigm relies on a predefined stage order.
This rigidity limits adaptivity to different query demands and intermediate evidence.
For straightforward questions, the fixed procedure may also introduce unnecessary steps, extra noise, and a higher risk of hallucination~\cite{cuadron2025danger, elliesql}.
For complex cases, once an early stage misses key schemas or values, later stages may proceed under a flawed premise, leading to error propagation.
Although intermediate artifacts such as filtered schemas and execution feedback can provide useful signals, a fixed pipeline can only utilize them for later stages rather than reconsider earlier decisions.

A more flexible direction is to coordinate \nlsql through workflow orchestration.
Rather than applying the same workflow to every question, orchestration selectively composes specialized modules into a query-specific workflow~\cite{flowreasoner, aflow, ruan2026aorchestra}.
This separates \emph{subtask reasoning} from \emph{process organization}: specialized modules handle concrete \nlsql subtasks, while an orchestration policy decides which to invoke and how to combine them~\cite{nielsen2026learning, zhang2026orchestrao1, deepeye_demo}.

\stitle{Attempts at \nlsql Orchestration.}
Some recent methods have started to explore orchestration for Text-to-SQL. A typical strategy is \textit{plan-then-execute orchestration}~\cite{beyond_static_pipelines} as illustrated by Figure~\ref{fig:nlsql_compare}(b): the system first plans a complete workflow for the input query, and then invokes action modules in the order specified by the planned workflow.
Compared with fixed pipelines, plan-then-execute orchestration provides better flexibility by producing a query-specific workflow.
However, as shown in Figure~\ref{fig:nlsql_compare}(b), once the workflow execution begins, its procedure becomes fixed again.
As a result, the system is still limited in adjusting later decisions based on intermediate artifacts or execution feedback.

These limitations suggest that planning a complete workflow before execution is not sufficient.
A more desirable approach is to leverage intermediate results for \emph{step-wise orchestration} as illustrated in Figure~\ref{fig:nlsql_compare}(c): instead of committing to the entire workflow upfront, the system can use intermediate artifacts and execution feedback to decide the next action accordingly.
This can enable the workflow to better adapt to the solving demand and evolving reasoning process of each query.

\stitle{Challenges.}
However, learning such a step-wise orchestration policy is non-trivial, mainly due to the following challenges.

\textbf{(C1) Heterogeneous Workflow Demands Across Queries.}
Different queries can require diverse workflow compositions.
Some queries may be resolved through direct \sql generation, whereas others may require additional evidence-gathering, refinement, or iterative generation as intermediate feedback emerges. 
Therefore, effective orchestration requires both a composable decision space and a policy that adapts its orchestration decisions to each query's evolving reasoning process.

\textbf{(C2) Scarce Workflow Data for Orchestration Learning.}
Learning such adaptive orchestration requires large-scale, high-quality, and diverse \nlsql workflow data that covers different queries and databases, workflow compositions, or decision patterns and reasoning trajectories. However, such data is rarely available and impractical to obtain through manual annotation. This raises the challenge of deriving workflow-level supervision without relying on human-crafted workflows.

\textbf{(C3) Large, Non-Unique, and Noisy Workflow Space.}
The workflow space is combinatorial and inherently diverse. A question may admit multiple valid workflows, and no single canonical workflow can generally serve as the ground truth. 
Moreover, a workflow that reaches the correct answer is not necessarily reliable, since its success may be accidental or unstable.
Effective orchestration learning therefore requires identifying high-quality workflows while preserving useful workflow diversity.

\stitle{Our Proposal.}
To systematically address these challenges, we propose \sys, a step-wise orchestration learning framework for \nlsql. 
\textbf{To address C1,} \sys formulates typical \nlsql subtasks as specialized action modules in an action space, and trains a policy model to make step-wise orchestration decisions based on the current solving context. This allows the system to compose workflows with flexible action combinations for various query demands. 
\textbf{To address C2,} \sys introduces \emph{Search-to-Policy Learning}, where \emph{MCTS-based Workflow Exploration} treats workflow construction as a tree-structured search over the action space. By balancing broad exploration with execution-guided selection, it automatically discovers diverse promising workflows, avoiding the need for human-crafting. 
\textbf{To address C3,} \sys further estimates the stability of explored workflows and curates reliable yet diverse trajectories for policy learning. The policy model is first trained with \emph{Stability-weighted Supervised Fine-tuning} to absorb robust orchestration patterns, and is then enhanced through \emph{Curriculum Reinforcement Learning}. 

In this way, broad offline search is converted into a deployable policy that can make efficient step-wise orchestration decisions at inference time to adapt to the different demands and evolving reasoning processes for diverse queries.

\stitle{Contributions.}
We make the following contributions:

\textbf{(S1) A Step-wise Orchestration Learning Framework.}
We propose \sys, a step-wise orchestration learning framework for \nlsql that separates process organization from subtask execution. 
Instead of planning and fixing a complete workflow upfront, \sys learns to dynamically orchestrate next action based on intermediate artifacts and feedback, enabling query-specific workflows with adaptive depths and module compositions.

\textbf{(S2) Search-to-Policy Learning for Adaptive Orchestration.}
We introduce a \textit{Search-to-Policy Learning} mechanism to train the orchestration policy. 
We first perform MCTS-based Workflow Exploration to discover high-quality and diverse candidate trajectories, then curate reliable workflows through stability estimation, training the policy model with Stability-weighted Supervised Fine-tuning to prioritize robust orchestration patterns while preserving diversity, followed by Curriculum Reinforcement Learning for further enhancement.
This transforms offline workflow search into an inference-time policy for adaptive orchestration.

\textbf{(S3) Extensive Experiments and Empirical Insights.}
\sys achieves superior execution accuracy, reaching 73.2\% EX on BIRD-Dev by training only a much smaller orchestration policy to coordinate frozen larger action models, outperforming prior methods that directly train comparable or larger \nlsql backbones.
Additionally, \sys demonstrates strong generalization across out-of-distribution datasets without specific training. 
Further analyses also reveal how \sys adapts workflow orchestration to diverse queries.

\vspace{-0.0em}

\section{Preliminaries}\label{sec:preliminary}
\vspace{-0.0em}

\subsection{\nlsql}\label{sec:preliminary_nlsql}

The \nlsql task aims to translate a natural language question into an executable \sql query over a relational database. 
Formally, given a natural language question \(q\) and a database \(\mathcal{D}\), a \nlsql system generates an \sql query \(y\). 
Executing \(y\) on \(\mathcal{D}\) produces the query result used to answer \(q\), denoted as \(\mathrm{Exec}_{\mathcal{D}}(y)\).

\nlsql is primarily evaluated with execution accuracy (EX), which measures whether the predicted \sql query produces the same execution result as the ground-truth query~\cite{bird}. 
Let \(y^\ast\) denote the ground-truth \sql. 
A predicted query \(y\) is considered correct if:
\[\label{eq:correct_sql}
\mathrm{Exec}_{\mathcal{D}}(y) = \mathrm{Exec}_{\mathcal{D}}(y^\ast).
\]

\subsection{Workflow Orchestration for \nlsql}

To handle the coordinated reasoning required by real-world \nlsql, workflow orchestration provides a flexible formulation that organizes specialized action modules into a workflow. 
Formally, let \(\mathcal{A}=\{a^{(1)},a^{(2)},\ldots,a^{(K)}\}\) denote an action space.
A workflow is a finite action sequence:
\[
\tau = (a_1, a_2, \ldots, a_T), \quad a_t \in \mathcal{A}.
\]
Executing a workflow on \((q,\mathcal{D})\) further induces a trajectory that records the orchestration states and intermediate action outputs:
\[\label{eq:traj}
\xi = (s_0, a_1, o_1, s_1, \ldots, a_T, o_T, s_T), 
\quad o_t = a_t(s_{t-1}, \mathcal{D}),
\]
where \(s_0\) is the initial orchestration state induced by \((q,\mathcal{D})\), \(o_t\) is the output of action \(a_t\), and \(s_t\) is the updated state after incorporating \(o_t\).
The workflow can therefore be viewed as the action-level abstraction extracted from a trajectory.
In \nlsql, the final prediction \(y\) is obtained from the terminal state.

\stitle{Plan-then-Execute Orchestration.}
One orchestration strategy is to first use an orchestration policy \(\pi\) to generate a complete workflow, as shown in Figure~\ref{fig:nlsql_compare}(b):
\[
\tau = \pi(q, \mathcal{D}).
\]
The actions in \(\tau\) are then invoked in order. 
Compared with a universal fixed pipeline, plan-then-execute orchestration provides a query-specific workflow. 
However, since the entire workflow is determined before execution, the action sequence is not revisable according to intermediate evidence produced during execution.

\stitle{Step-wise Orchestration.}
In contrast, step-wise orchestration applies the orchestration policy during execution, as shown in Figure~\ref{fig:nlsql_compare}(c).
At each step \(t\), the policy selects the next action according to the current orchestration state, \ie
\[
a_t=\pi(s_{t-1}), \quad \tau_\pi=(a_1,a_2,\ldots,a_T).
\]
After the selected action is invoked, its output is incorporated into the orchestration state, which forms the context for the next decision.
Since each state incorporates previous actions and their outputs, the next decision can condition on updated evidence, making the workflow adaptive to the evolving solving context of each query.
In this paper, we focus on learning such a step-wise workflow orchestration policy for \nlsql.
\section{\sys Overview}\label{sec:overview}


\begin{figure*}[t!]
    \centering
    \includegraphics[width=1.0\linewidth]{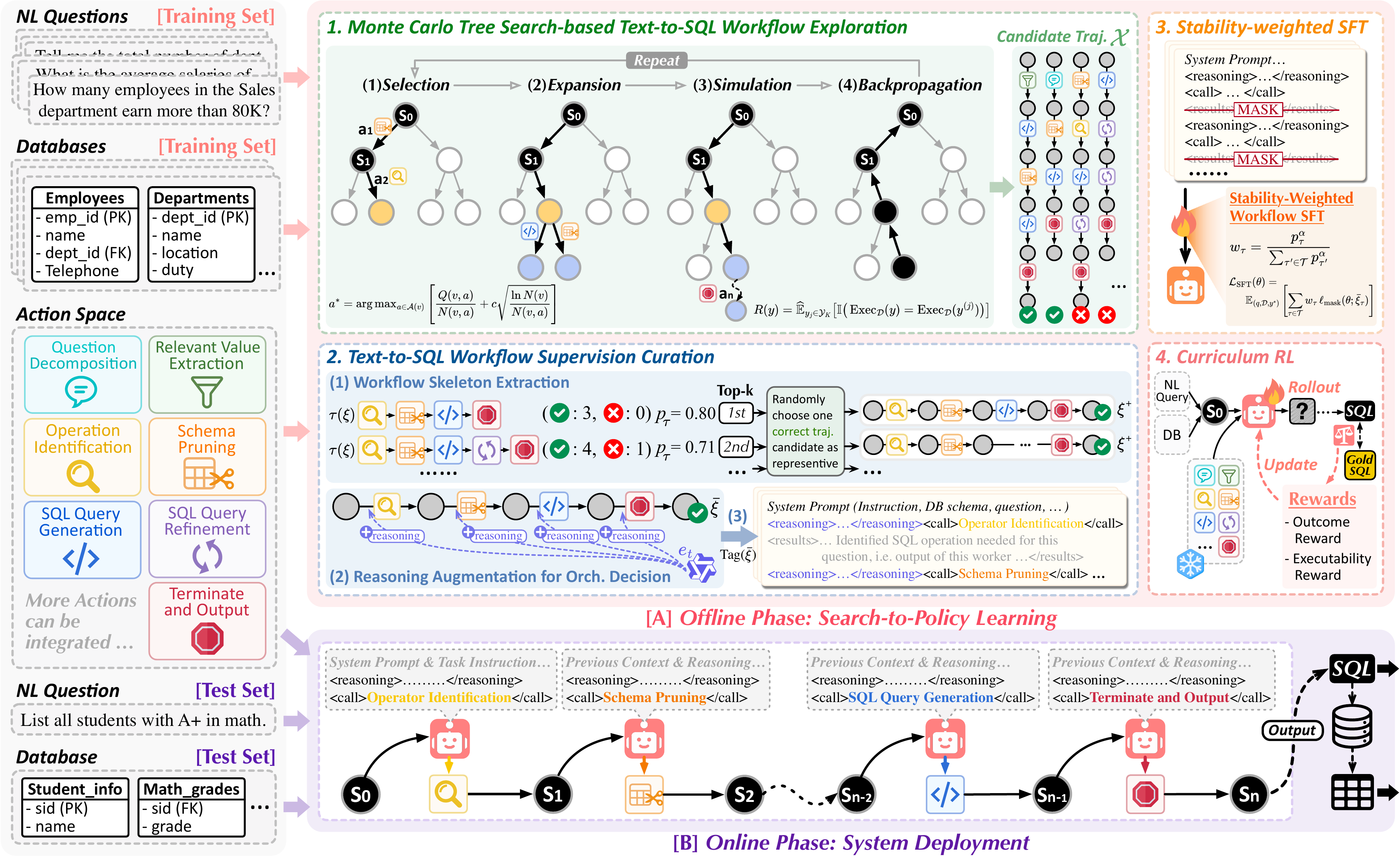}
    \vspace{-2.0em}
    \caption{The framework overview of \sys.}
    \label{fig:framework}
    \vspace{-1.0em}
\end{figure*}

Figure~\ref{fig:framework} illustrates the overall framework of \sys, which consists of two phases: offline \textit{Search-to-Policy Training} and online \textit{Deployment}.
In the offline phase, \sys uses MCTS to explore \nlsql workflows, curates workflow supervision, and trains a policy model with stability-weighted supervision and reinforcement learning.
In the online phase, the learned policy model coordinates action modules step by step to solve each input query.
This design separates \emph{process organization} from \emph{subtask reasoning}: the trainable policy learns how to orchestrate effective \nlsql workflow, while frozen LLM-driven specialized action modules carry out concrete subtasks in the workflow.

\stitle{Offline: Search-to-Policy Training.}
As illustrated in Figure~\ref{fig:framework}-A, the goal of the offline phase is to learn a step-wise orchestration policy from automatically constructed workflow supervision.

\textit{(1) MCTS-based Workflow Exploration.}
\sys first performs Monte Carlo Tree Search over the action space.
Each node represents an orchestration state, and each edge corresponds to an action selection.
The search process broadly explores different action compositions and workflow depths, while using execution outcomes to guide the discovery of promising workflows.
This produces a diverse set of candidate workflows for each training instance, covering a wide range of potential solving paths.

\textit{(2) Workflow Supervision Curation.}
The searched workflows are diverse but may contain unstable or accidental successes.
Therefore, \sys further processes and curates workflow supervision before training.
Specifically, it extracts workflow skeletons, \ie action sequences, from candidate trajectories, groups with the same skeleton, and estimates the stability of each workflow according to its success and failure statistics.
Then, \sys selects top-ranked workflow skeletons and converts their representative trajectories into step-level instruction-tuning formatted data, which contains the current orchestration context, its next action, and the corresponding stability score.
This step filters unreliable workflows while preserving diverse orchestration patterns.

\textit{(3) Stability-weighted Supervised Fine-tuning.}
Based on the curated training samples, \sys introduces Stability-weighted Supervised Fine-tuning to train the policy model in step 3.
Different from treating all selected samples equally, this training objective assigns larger weights to samples with higher stability scores.
In this way, the policy is encouraged to learn more robust orchestration patterns from explored workflows, while still retaining alternative valid solving strategies.

\textit{(4) Curriculum Reinforcement Learning.}
In step 4, \sys further enhances the policy through curriculum reinforcement learning.
The curriculum emphasizes challenging queries that are less effectively solved during MCTS-based exploration.
For these instances, the policy rolls out to orchestrate workflows and receives rewards based on execution-oriented feedback.
This further calibrates the policy for complex cases, especially in deciding when to collect additional evidence or regenerate a new query iteratively.

\stitle{Online: System Deployment.}
After training, \sys is deployed as a step-wise \nlsql workflow orchestration system.
Given an unseen question \(q\) and database \(\mathcal{D}\), 
the learned policy model iteratively selects an action based on the current orchestration state; the system then automatically parses to invoke the corresponding action module and updates the state with its output.
This process continues until the termination is determined, after which \sys outputs the current \sql query as the final prediction.
Thus, the workflow is constructed during execution and adapted to the evolving solving context of each query.

\section{The Design Details of \sys} \label{sec:details}

\subsection{Action Space}\label{sec:action_space}

\sys separates workflow orchestration from subtask reasoning.
To support flexible workflow composition, we expose typical \nlsql capabilities as callable action modules. 
Following prior practice~\cite{alphasql, wang2025squrve}, we implement seven actions for typical subtasks in \nlsql to instantiate the action space.
Formally,
\begin{equation}
\mathcal{A} =
\{
a_{\mathrm{dec}},
a_{\mathrm{val}},
a_{\mathrm{op}},
a_{\mathrm{sch}},
a_{\mathrm{gen}},
a_{\mathrm{ref}},
a_{\mathrm{term}}
\},
\label{eq:action-space}
\end{equation}
where the first six actions invoke frozen LLM-driven subtask modules, and \(a_{\mathrm{term}}\) terminates the workflow and returns predicted \sql.
Given an orchestration state \(s_{t-1}\), invoking an action \(a_t \in \mathcal{A}\) produces an action-specific output \(o_t\), which is incorporated into the subsequent state \(s_t\) following Eq.~\eqref{eq:traj}.

Importantly, Eq.~\eqref{eq:action-space} defines available capabilities rather than a predefined execution order.
Conditioned on the current intermediate artifacts and feedback, the policy may skip unnecessary actions, gather additional evidence, revisit an earlier action, regenerate a candidate query, refine an existing query, or terminate when obtaining a sufficient \sql candidate.
Figure~\ref{fig:framework} summarizes the action space, and we describe each action below.

\ptitle{Question Decomposition (\(a_{\mathrm{dec}}\)).}
\nlq questions may compactly express multiple constraints or dependencies~\cite{ma2024plug}.
This action rewrites the original question into fine-grained requirements while preserving semantics, making the underlying sub-intents easier to consume in later reasoning.
For example, ``How many employees in the Sales department earn more than 80K?'' can be decomposed into identifying Sales employees, filtering salaries above 80K, and counting the remaining employees.

\ptitle{Relevant Value Extraction (\(a_{\mathrm{val}}\)).}
\nlq questions often contain values that should be grounded into \sql predicates.
Given the question and database information, this action infers condition values likely needed for \sql construction and provides predicate-level hints.
For example, for the Sales-department question above, it may identify ``Sales'' and ``80K'' as relevant condition values.

\ptitle{Operation Identification (\(a_{\mathrm{op}}\)).}
Questions may imply \sql operations such as aggregation, grouping, ordering, comparison, deduplication, temporal transformation, scalar functions, or nested queries.
This action records operation-level requirements as intermediate evidence to reduce the
burden on later analysis.
For example, it may identify \texttt{AVG(salary)} for an average-salary question, or a ranking pattern like \texttt{ORDER BY salary DESC LIMIT 1}.

\ptitle{Schema Pruning (\(a_{\mathrm{sch}}\)).}
Large databases often contain many irrelevant tables and columns.
Based on the question and accumulated evidence, this action selects a compact schema subset that preserves the tables, columns, and relational paths needed for the current query.
Because it remains callable throughout the workflow, the policy may invoke it again when later feedback suggests that essential schema elements may be omitted.

\ptitle{SQL Query Generation (\(a_{\mathrm{gen}}\)).}
This action synthesizes a complete \sql candidate from the current orchestration state.
Following DAIL-SQL~\cite{dailsql}, \sys adopts few-shot in-context learning, and the LLM is prompted with the question, schema, few-shot examples, and accumulated evidence.
The generated query is executed on the database, and the feedback is incorporated into the updated state, allowing generation to be invoked again after new values, schema elements, or operation cues are obtained.

\ptitle{SQL Query Refinement (\(a_{\mathrm{ref}}\)).}
When an existing \sql candidate requires targeted correction, \sys can apply this action following the strategy of DeepEye-SQL~\cite{deepeyesql}.
The candidate is checked by a fail-fast toolchain covering syntax and execution validation, logical verification, and query-quality checking.
If a defect is detected, the diagnostic signal is converted into an explicit revision instruction for the LLM to produce a corrected candidate.

\ptitle{Terminate and Output (\(a_{\mathrm{term}}\)).}
This terminal action ends the workflow and returns the latest \sql candidate as the prediction.
Allowing the policy to select \(a_{\mathrm{term}}\) adaptively avoids redundant modules for straightforward questions while preserving the option to invest more reasoning steps when evidence remains insufficient.

The action space is intentionally modular.
Additional \nlsql capabilities can be incorporated as new actions without changing the overall orchestration formulation.
In the next subsection, we describe how \sys explores compositions of these actions through Monte Carlo Tree Search to construct diverse candidate workflows for policy learning.

\subsection{MCTS-based Workflow Exploration}\label{sec:mcts-exploration}

\begin{algorithm}[t]
\caption{MCTS-based Workflow Exploration}
\label{alg:mcts-exploration}
\begin{algorithmic}[1]
\small
\Require Question \(q\), database \(\mathcal{D}\), action space \(\mathcal{A}\), rollout budget \(N_{\mathrm{rollout}}\), expansion samples \(N_{\mathrm{exp}}\), reward samples \(K\), maximum depth \(T_{\max}\)
\Ensure Candidate trajectory set \(\mathcal{X}(q,\mathcal{D})\)
\State \(s_0\leftarrow\operatorname{InitState}(q,\mathcal{D})\)
\State Initialize \(\Psi=(V,E)\) with root \(v_0\) where \(s(v_0)=s_0\); \(\mathcal{X}(q,\mathcal{D})\leftarrow\emptyset\)
\For{\(i=1,\ldots,N_{\mathrm{rollout}}\)}
    \State \(v\leftarrow v_0\)
    \While{\(v\) is non-terminal and fully expanded}
        \State \(a\leftarrow\operatorname{UCTSelect}(v)\), \(v\leftarrow\operatorname{Child}(v,a)\) \hfill$\triangleright$ Selection, Eq.~\eqref{eq:mcts-uct}
    \EndWhile
    \If{\(v\) is non-terminal and \(\operatorname{Depth}(v)<T_{\max}\)}
        \State \(\mathcal{C}\leftarrow\operatorname{Expand}(v,\mathcal{A},N_{\mathrm{exp}},\mathcal{D})\), \(v\leftarrow\operatorname{Sample}(\mathcal{C})\) \hfill$\triangleright$ Expansion
    \EndIf
    \While{\(v\) is non-terminal and \(\operatorname{Depth}(v)<T_{\max}\)}
        \State \(\mathcal{C}\leftarrow\operatorname{Expand}(v,\mathcal{A},N_{\mathrm{exp}},\mathcal{D})\), \(v\leftarrow\operatorname{Sample}(\mathcal{C})\) \hfill$\triangleright$ Simulation
    \EndWhile
    \State \textbf{if} \(v\) is non-terminal \textbf{then} \(v\leftarrow\operatorname{ForceTerminate}(v)\) 
    \State \(\xi\leftarrow\operatorname{Trace}(v_0,v)\), \(y\leftarrow\operatorname{ExtractSQL}(s(v))\)
    \State \(r\leftarrow R(y;\mathcal{D},K)\) \hfill$\triangleright$ Eq.~\eqref{eq:mcts-reward}
    \State \(\operatorname{Backpropagate}(\xi,r)\) \hfill$\triangleright$ Backpropagation
    \State \(\mathcal{X}(q,\mathcal{D})\leftarrow\mathcal{X}(q,\mathcal{D})\cup\{\xi\}\)
\EndFor
\State \Return \(\mathcal{X}(q,\mathcal{D})\)
\end{algorithmic}
\end{algorithm}

The action space introduced in Section~\ref{sec:action_space} enables flexible workflow composition, but also induces a large combinatorial search space. Manually specifying high-quality workflows for different questions is therefore impractical, while directly sampling trajectories from an LLM may cover only a limited set of orchestration patterns. To automatically discover diverse orchestration patterns, \sys performs Monte Carlo Tree Search (MCTS)~\cite{browne2012mcts_survey, kocsis2006bandit} over the action space for each training instance \((q,\mathcal{D})\).

\stitle{Tree-structured Workflow Search.}
We construct a search tree \(\Psi=(V,E)\) rooted at the initial orchestration state \(s_0\).
Each node \(v \in V\) stores an orchestration state \(s(v)\) reached after a sequence of action invocations, and each edge corresponds to an action-induced state transition.
Specifically, invoking action \(a_t\) at state \(s_{t-1}\) produces an intermediate output \(o_t\) and an updated state \(s_t\).
A root-to-terminal path therefore induces a trajectory
\(
\xi=(s_0,a_1,o_1,s_1,\ldots,a_T,o_T,s_T),
\)
whose action-only projection \(\tau(\xi)=(a_1,\ldots,a_T)\) is the corresponding workflow.
Trajectories retain concrete intermediate artifacts, while workflows abstract action-composition patterns.

Algorithm~\ref{alg:mcts-exploration} summarizes the exploration procedure.
As illustrated in Figure~\ref{fig:framework}-A(1), each rollout consists of selection, expansion, simulation, and backpropagation.

\stitle{Selection.}
Starting from the root, MCTS traverses the current tree until it reaches a terminal state or a node with unexplored children.
At each visited node \(v\), the next action is selected by the Upper Confidence Bound for Trees (UCT) criterion~\cite{kocsis2006bandit}:
\begin{equation}
a^{*} =
\argmax_{a \in \mathcal{A}(v)}
\left[
\frac{Q(v,a)}{N(v,a)}
+
c \sqrt{
\frac{\ln N(v)}{N(v,a)}
}
\right],
\label{eq:mcts-uct}
\end{equation}
where \(\mathcal{A}(v)\) is applicable action set at node \(v\);
\(Q(v,a)\) and \(N(v,a)\) denote accumulated reward and visit count of edge \((v,a)\);
\(N(v)\) is the visit count of \(v\); and \(c\) controls the exploration--exploitation trade-off.
Unvisited actions are prioritized before applying Eq~\eqref{eq:mcts-uct}.

\stitle{Expansion.}
When traversal reaches a non-terminal frontier node, \sys expands the tree by invoking applicable action modules.
To preserve exploration diversity, the same action may be sampled multiple times with a non-zero temperature, producing alternative child states under the same preceding context.
Each child stores the updated orchestration state and the newly generated intermediate artifact.

\stitle{Simulation.}
Starting from an expanded child, \sys continues sampling and invoking actions until \texttt{Terminate and Output} is selected.
Because each state incorporates accumulated artifacts and feedback, later actions operate on an evolving solving context.
When a rollout reaches a terminal state, \sys extracts its \sql candidate \(y\) and evaluates it with an execution-consistency reward.
Following existing research~\cite{alphasql, cscsql}, we sample additional \sql candidates
\(\mathcal{Y}_K=\{y^{(1)},\ldots,y^{(K)}\}\)
from the current SQL-generation context and compute
\begin{equation}
R(y) = \widehat{\mathbb{E}}_{y^{(j)} \in \mathcal{Y}_K} \big[\mathbb{I}\big(\operatorname{Exec}_{\mathcal{D}}(y) = \operatorname{Exec}_{\mathcal{D}}(y^{(j)})\big)\big].
\label{eq:mcts-reward}
\end{equation}
If \(y\) is non-executable, we set \(R(y)=0\).
This reward assigns higher scores to candidates whose execution results are reproduced consistently across multiple generations, providing a lightweight execution-oriented signal for guiding exploration.

\stitle{Backpropagation.}
The reward is propagated from the terminal node back to root, updating visit counts and accumulated rewards of the edges along the explored path.
Repeating this procedure allocates more rollouts to promising regions while still exploring alternative action compositions.
Meanwhile, \sys records completed trajectories as candidates for subsequent supervision curation.
After \(N_{\mathrm{rollout}}\) iterations, the collected trajectory set is
\begin{equation}
    \mathcal{X}(q,\mathcal{D})
    =
    \left\{
        \xi^{(1)}, \xi^{(2)}, \ldots, \xi^{(M)}
    \right\}.
    \label{eq:candidate-trajectories}
\end{equation}

\subsection{Workflow Supervision Curation}\label{sec:curation}

MCTS-based exploration produces diverse candidate trajectories for each training instance.
However, these trajectories cannot be directly used as policy supervision.
The explored trajectories only record action calls and action outputs, without explicit reasoning for orchestration decisions, and some successful trajectories may be accidental.
Therefore, \sys curates the explored trajectories into reliable, reasoning-augmented, and instruction-tuning-ready supervision, as shown in Figure~\ref{fig:framework}-A(2) and detailed in Algorithm~\ref{alg:curation}.

\begin{figure}[t!]
    \centering
    \vspace{-1.0em}
    \includegraphics[width=1.0\columnwidth]{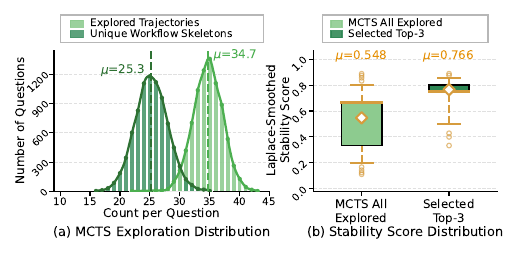}
    \vspace{-2.75em}
    \caption{Statistics of workflow exploration and curation.}
    \label{fig:curation_stats}
    \vspace{-1.25em}
\end{figure}

\stitle{Workflow Skeleton Extraction.}
Given a MCTS explored trajectory \(\xi\), we first extract its workflow skeleton (\ie workflow \(\tau\)) by keeping only its action sequence:
\(
\tau(\xi)=(a_1,a_2,\ldots,a_T).
\)
Different trajectories with the same workflow are grouped together:
\[\label{eq:group}
\mathcal{X}_\tau=\{\xi\in\mathcal{X}(q,\mathcal{D})\mid \tau(\xi)=\tau\},
\]
where \(\mathcal{X}(q,\mathcal{D})\) denotes the set of candidate trajectories explored for the training instance.
For each workflow \(\tau\), we count the number of successful trajectories \(s_\tau\) and the total number of trajectories \(n_\tau=|\mathcal{X}_\tau|\), where success is determined by execution correctness. 
Figure~\ref{fig:curation_stats}(a) shows that MCTS explores a broad set of trajectories with diverse workflow skeletons for each question.

\stitle{Workflow Stability Estimation.}
Since the combinatorial space is large, most workflows are observed only once or a limited number of times. 
To account for sparse observations, \sys defines each workflow's stability score as the posterior predictive probability of success under a Bernoulli model with a uniform $\operatorname{Beta}(1,1)$ prior, as given by Laplace's rule of succession~\cite{zabell1989rule}:
\[\label{eq:laplace}
\hat{p}_\tau = \frac{s_\tau+1}{n_\tau+2},
\]
The resulting \(\hat{p}_\tau\) serves as a Laplace-smoothed workflow stability score.
It assigns higher scores to workflows that repeatedly lead to successful executions, while avoiding extreme estimates for rarely observed workflows.
For each training instance, \sys ranks all workflows with \(s_\tau>0\) by \(\hat{p}_\tau\) to select the top-\(k\) workflows, and randomly samples one successful trajectory from each selected group as the representative trajectory \(\xi^+\).
Since grouped trajectories share the same high-level orchestration decisions, retaining one representative per workflow reduces redundant supervision and exploration-frequency bias.
This keeps multiple valid workflow patterns for the same question, rather than forcing a single canonical workflow.
Figure~\ref{fig:curation_stats}(b) shows that the selected top-\(3\) workflows have clearly higher stability scores among all explored workflows.

\stitle{Reasoning Augmentation for Orchestration Decisions.}
The representative trajectory \(\xi^+\) contains the selected actions and their outputs, but the policy model should also learn why an action is selected under the current orchestration context.
Thus, we augment each representative trajectory with step-level orchestration reasoning.
For each step \(t\), we prompt a deployed LLM to generate the orchestration reasoning \(e_t=\mathrm{LLM}(q,\mathcal{D},\bar{\xi}_{<t},a_t,o_t)\) for invoking \(a_t\), where \(\bar{\xi}_{<t}\) is the augmented trajectory prefix before step \(t\), and \(a_t,o_t\) are the selected current action and its observed output.
The reasoning is inserted before the corresponding action call, yielding:
\[
\bar{\xi}
=
(s_0,e_1,a_1,o_1,s_1,\ldots,e_T,a_T,o_T,s_T).
\]
This step-wise augmentation is more controllable than generating reasoning for the whole trajectory at once, because the LLM only needs to justify a fixed action under a fixed context and observed result.
Since the trajectory has already been selected from stable successful candidates, the generated reasoning is grounded in high-quality orchestration behavior.

\begin{algorithm}[t]
\caption{Workflow Supervision Curation}
\label{alg:curation}
\begin{algorithmic}[1]
\small
\Require Training set \(\mathcal{I}\), MCTS explored trajectories \(\mathcal{X}(q,\mathcal{D})\), top-\(k\)
\Ensure Curated supervision set \(\mathcal{C}\)
\State \(\mathcal{C}\leftarrow\emptyset\)
\ForAll{\((q,\mathcal{D},y^\ast)\in\mathcal{I}\)}
    \State Extract \(\tau(\xi)=(a_1,\ldots,a_T)\) for each \(\xi\in\mathcal{X}(q,\mathcal{D})\)
    \State \(\mathcal{X}_\tau\leftarrow\{\xi\in\mathcal{X}(q,\mathcal{D})\mid \tau(\xi)=\tau\}\) for each distinct \(\tau\) \hfill$\triangleright$ Eq.~\eqref{eq:group}
    \ForAll{workflow group \(\mathcal{X}_\tau\)}
        \State \(s_\tau\leftarrow|\{\xi\in\mathcal{X}_\tau:\xi\text{ is successful}\}|,\ n_\tau\leftarrow|\mathcal{X}_\tau|\) 
        \State \(\hat{p}_\tau\leftarrow\dfrac{s_\tau+1}{n_\tau+2}\) \Comment{Laplace-smoothed stability score} \hfill$\triangleright$ Eq.~\eqref{eq:laplace}
    \EndFor
    \State \(\mathcal{T}\leftarrow\text{top-}{k}(\{\tau\mid s_\tau>0\};\hat{p}_\tau)\) \Comment{Select top-\(k\) workflows}
    \ForAll{\(\tau\in\mathcal{T}\)}
        \Statex \quad\quad\quad \Comment{Randomly select a successful representative trajectory}
        \State \(\xi^+=(s_0,a_1,o_1,\ldots,a_T,o_T,s_T)\in\mathcal{X}_\tau\)
        \State Initialize \(\bar{\xi}\leftarrow(s_0)\)
        \For{\(t=1,\ldots,T\)}
            \State \(e_t\leftarrow\mathrm{LLM}(q,\mathcal{D},\bar{\xi}_{<t},a_t,o_t)\), \(\bar{\xi}\leftarrow\bar{\xi}\oplus(e_t,a_t,o_t,s_t)\)
        \EndFor
        \State \(\mathcal{C}\leftarrow\mathcal{C}\cup\{(\operatorname{Tag}(\bar{\xi}),\hat{p}_\tau)\}\)
    \EndFor
\EndFor
\State \Return \(\mathcal{C}\)
\end{algorithmic}
\end{algorithm}

Finally, as shown in Figure~\ref{fig:curation_format}, we serialize each augmented trajectory into a tagged instruction-tuning format, where \texttt{<reasoning>} and \texttt{<call>} are supervised as policy-generated content, and \texttt{<results>} is the output of the corresponding action.

\begin{figure}[h]
    \centering
    \vspace{-0.5em}
    \setlength{\fboxsep}{2pt}
    \fbox{
    \begin{minipage}{1.005\columnwidth}
    \small
    \texttt{System Prompt: } \(q, \mathcal{D}, \mathcal{A}, \text{instruction}, \ldots\)\\
    \texttt{<reasoning>}\(e_1\)\texttt{</reasoning><call>}\(a_1\)\texttt{</call><results>}\(o_1\)\texttt{</results>}\\
    \(\cdots\)\\
    \texttt{<reasoning>}\(e_T\)\texttt{</reasoning><call>}\(a_T\)\texttt{</call><results>}\(o_T\)\texttt{</results>}
    \end{minipage}
    }
    \vspace{-1.0em}
    \caption{Instruction-tuning format for subsequent training.}
    \label{fig:curation_format}
    \vspace{-1em}
\end{figure}

\subsection{Stability-weighted Supervised Fine-tuning}\label{sec:sft}

After workflow supervision curation, each selected sample consists of a tagged augmented trajectory \(\operatorname{Tag}(\bar{\xi})\) and its workflow stability score \(\hat{p}_\tau\).
This stage introduces a \textit{Stability-weighted Supervised Fine-tuning} to train the policy to learn from robust orchestration patterns from selected workflows, while still preserving diversity from alternative valid solving strategies.

\stitle{Action Result Masking.}
In \sys, the training objective focuses on learning orchestration for \nlsql workflow.
Thus, different from fine-tuning \nlsql generators, \sys only updates the compact policy model.

As mentioned in Section~\ref{sec:curation}, the tagged instruction-tuning formatted data contains both policy-generated content and action-generated observations.
Since \texttt{<results>} is not produced by the policy, forcing the policy to predict it would introduce irrelevant supervision and bias the model from learning workflow orchestration~\cite{ma2026empowering}.
Therefore, \sys keeps \texttt{<results>} in the context for subsequent decisions, but masks it out from the training loss~\cite{ma2026empowering, feng2026retool, zhang2025tool}.
Formally, let \(\operatorname{Tag}(\bar{\xi})\) be the serialized token sequence.
We define a tag-aware binary mask \(m_i\) as:
\[\label{eq:tagged_sequence}
\operatorname{Tag}(\bar{\xi})=(z_1,z_2,\ldots,z_N),
\]
\[
m_i =
\begin{cases}
0, & z_i \in \texttt{<results>}\cdots\texttt{</results>},\\
1, & \text{otherwise}.
\end{cases}
\]
In practice, the system prompt and other non-response tokens are also excluded following the standard instruction-tuning setting.
The result-masked loss is:
\[
\ell_{\mathrm{mask}}(\theta;\bar{\xi})
=
-\frac{1}{\sum_{i=1}^{N}m_i}
\sum_{i=1}^{N}
m_i \log P_\theta(z_i\mid z_{<i}).
\]
This normalization prevents longer trajectories or longer action outputs from dominating the optimization.
Meanwhile, the masked results remain visible in the context, allowing the policy to learn state-conditioned decisions from evolving intermediate evidence.

\stitle{Stability-weighted Objective.}
Although the curated trajectories are selected from successful candidates, they are not equally reliable.
A workflow that succeeds repeatedly during MCTS-based exploration provides stronger evidence of robust orchestration behavior than one that succeeds only occasionally.
Therefore, \sys uses the Laplace-smoothed workflow stability score \(\hat{p}_\tau\) from Eq.~\eqref{eq:laplace} to weight the supervised learning objective.

For the selected trajectories of the same question, \sys converts stability scores into normalized training weights.
This avoids treating \(\hat{p}_\tau\) as an absolute loss scale and instead uses it to define the relative importance of different valid workflows for the same question.
Specifically, we adopt power normalization:
\[
w_\tau
=
\frac{\hat{p}_\tau^{\alpha}}
{\sum_{\tau'\in\mathcal{T}}p_{\tau'}^{\alpha}},
\]
where \(\mathcal{T}\) denotes the selected top-\(k\) workflows for a certain question.
The exponent \(\alpha\) controls the strength of stability preference.
Following prior work on distribution sharpening~\cite{berthelot2019mixmatch}, we set \(\alpha=2\).

The final Stability-weighted Supervised Fine-tuning objective is:
\[
\mathcal{L}_{\mathrm{SFT}}(\theta)
=
\mathbb{E}_{(q,\mathcal{D},y^\ast)\sim\mathcal{I}}
\left[
\sum_{\tau\in\mathcal{T}}
w_\tau\,
\ell_{\mathrm{mask}}(\theta;\bar{\xi}_\tau)
\right],
\]
where \(\bar{\xi}_\tau\) is the representative augmented trajectory of workflow \(\tau\).
This objective prioritizes stable orchestration patterns while still preserving multiple valid workflows for the same question.
As a result, the policy obtains a reliability-aware initialization for tagged interaction formatting and step-wise action decision, which is further enhanced by curriculum reinforcement learning.

\subsection{Curriculum Reinforcement Learning}\label{sec:rl}

Stability-weighted Supervised Fine-tuning trains the policy model to learn reliable orchestration patterns selected from MCTS exploration.
However, its supervision still comes from curated successful trajectories.
For difficult instances, successful trajectories can be sparse during exploration, making supervised signals less sufficient for learning how to recover from imperfect intermediate decisions.
Therefore, \sys further applies \textit{Curriculum Reinforcement Learning} to enhance the orchestration policy on challenging queries through self-rollouts and execution-oriented rewards.

\stitle{Curriculum Construction.}
We construct the reinforcement learning (RL) set according to the difficulty observed during MCTS-based exploration.
For each training instance \((q,\mathcal{D},y^\ast)\), let \(\mathcal{X}(q,\mathcal{D})\) denote the explored trajectory set, and let \(\mathcal{X}^{+}(q,\mathcal{D})\) denote the subset of successful ones obtaining correct predicted \sql.
We define the MCTS success ratio as:
\[
\rho(q,\mathcal{D})
=
\frac{|\mathcal{X}^{+}(q,\mathcal{D})|}
{|\mathcal{X}(q,\mathcal{D})|}.
\]
A lower \(\rho(q,\mathcal{D})\) indicates that successful trajectories are more sparsely yielded in the exploration, suggesting a more challenging query.
Given a threshold \(\gamma\), we construct the curriculum set as:
\[
\mathcal{I}_{\mathrm{RL}}
=
\{(q,\mathcal{D},y^\ast)\in\mathcal{I}
\mid
\rho(q,\mathcal{D}) < \gamma
\}.
\]
This curriculum directs the RL stage toward challenging instances that are less effectively solved during exploration, complementing curated supervision with further policy optimization.

\begin{figure}[t!]
    \centering
    \vspace{-0.5em}
    \includegraphics[width=1.0\columnwidth]{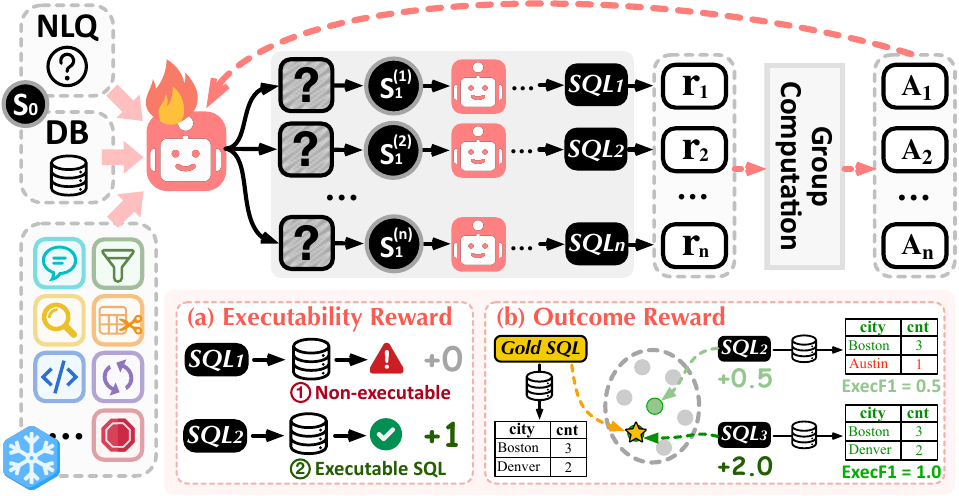}
    \vspace{-2.0em}
    \caption{Rollout and reward computation in curriculum reinforcement learning for \sys.}
    \label{fig:rl}
    \vspace{-1.5em}
\end{figure}

\stitle{Policy Rollout with GRPO.}
We adapt GRPO~\cite{shao2024deepseekmath, liu2026gdpo, nielsen2026learning, zhang2026orchestrao1} to workflow orchestration by optimizing the policy over grouped step-wise rollouts with execution-oriented rewards.
For each instance in \(\mathcal{I}_{\mathrm{RL}}\), the current policy performs step-wise rollout by interacting with the frozen action modules.
As shown in Figure~\ref{fig:rl}, starting from the initial state \(s_0\), the policy repeatedly generates orchestration reasoning and selects an action.
The corresponding action module is then invoked, with its output result appended to the context for the next decision.
The rollout stops when the policy decides to terminate and output or reaches the maximum step limit \(T_{\max}\), yielding a final predicted \sql query \(y_i\).
Each rollout is scored by an execution-oriented reward \(r_i\), which will be elaborated next.

Following GRPO, we sample a group of \(G\) rollouts for the same training instance:
\(
\{\xi_i,y_i,r_i\}_{i=1}^{G},
\)
where \(\xi_i\) is the rollout trajectory, \(y_i\) is its final predicted \sql, and \(r_i\) is the scalar reward computed from execution feedback.
GRPO estimates the advantage of each rollout by comparing it with other rollouts in the same group:
\[
\hat{A}_i
=
\frac{r_i-\mu_r}{\sigma_r+\delta},
\
\mu_r=\frac{1}{G}\sum_{j=1}^{G}r_j,
\]
where \(\sigma_r\) is the standard deviation of group rewards and \(\delta\) is a small constant for numerical stability.
This relative advantage avoids training an additional value model and encourages the policy to prefer workflows that perform better than other sampled workflows for the same question.

As in Eq.~\eqref{eq:tagged_sequence}, we serialize each rollout \(\xi_i\) into a token sequence \(z_{i,1:N_i}\).
Let \(\Omega_i\subseteq\{1,\ldots,N_i\}\) denote the index set of policy-generated tokens, including tokens in \texttt{<reasoning>} and \texttt{<call>}.
Let \(P_\theta\) denote the autoregressive token distribution induced by the policy model \(\pi_\theta\).
For each \(t\in\Omega_i\), we define the probability ratio:
\[
\eta_{i,t}
=
\frac{P_\theta(z_{i,t}\mid z_{i,<t})}
{P_{\theta_{\mathrm{old}}}(z_{i,t}\mid z_{i,<t})}.
\]
Therefore, our GRPO objective for \sys is:
\[
\begin{aligned}
\mathcal{J}_{\mathrm{GRPO}}(\theta)
=&
\frac{1}{G}
\sum_{i=1}^{G}
\frac{1}{|\Omega_i|}
\sum_{t\in\Omega_i}
\min\Big(
\eta_{i,t}\hat{A}_i,
\\
&
\operatorname{clip}(\eta_{i,t},1-\epsilon_c,1+\epsilon_c)\hat{A}_i
\Big)
-
\beta D_{\mathrm{KL}}(P_\theta\,\|\,P_{\mathrm{ref}})
\end{aligned}
\]
where \(P_{\theta_{\mathrm{old}}}\) is the policy used to generate the rollouts, \(P_{\mathrm{ref}}\) is the SFT-trained reference policy, and \(\epsilon_c\) and \(\beta\) control clipping and KL regularization.
As in SFT, only policy-generated tokens are optimized, while action results are treated as observations.

\stitle{Execution-oriented Reward.}
The reward is designed to encourage both executable \sql generation and semantically correct execution results.
Given a predicted \sql \(y_i\), \sys executes it on \(\mathcal{D}\) and compares its result with the execution result of the ground-truth \sql query \(y^\ast\).

\textit{(1) Executability Reward} encourages the policy to produce workflows that lead to valid \sql:
\[
r_{\mathrm{exe}}(y_i)
=
\mathbb{I}\left[
y_i \text{ executes on } \mathcal{D}
\text{ without error/timeout}
\right].
\]

\textit{(2) Outcome Reward} measures the quality of the final execution result.
Exact execution correctness remains the primary target, while partial result overlap can also be used to provide a more fine-grained signal for distinguishing near-correct predictions~\cite{guo2026dtbench}.
For this purpose, we employ ExecF1 over the execution result table of predicted and gold \sql.
Specifically, for executable \(y_i\), let \(\mathcal{V}_i\) and \(\mathcal{V}^\ast\) denote the multisets of normalized cell values in \(\mathrm{Exec}_{\mathcal{D}}(y_i)\) and \(\mathrm{Exec}_{\mathcal{D}}(y^\ast)\), respectively.
The numbers of matched, prediction-only, and gold-only values are:
\[
\mathrm{TP}=|\mathcal{V}_i\cap\mathcal{V}^\ast|,\ 
\mathrm{FP}=|\mathcal{V}_i\setminus\mathcal{V}^\ast|,\ 
\mathrm{FN}=|\mathcal{V}^\ast\setminus\mathcal{V}_i|,
\]
where repeated values are counted according to their multiplicities. As
\(
\mathrm{Prec}=\mathrm{TP}/(\mathrm{TP}+\mathrm{FP}),
\mathrm{Rec}=\mathrm{TP}/({\mathrm{TP}+\mathrm{FN}}),
\)
ExecF1 then denotes:
\[
\mathrm{ExecF1}(y_i,y^\ast)
=
\frac{2\cdot \mathrm{Prec}\cdot \mathrm{Rec}}
{\mathrm{Prec}+\mathrm{Rec}}.
\]
For non-executable \(y_i\), the outcome reward is also \(0\).
Otherwise, the outcome reward is:
\[
r_{\mathrm{out}}(y_i)
=
\begin{cases}
1+\mathrm{ExecF1}(y_i,y^\ast),
& \mathrm{Exec}_{\mathcal{D}}(y_i)
=
\mathrm{Exec}_{\mathcal{D}}(y^\ast),\\
\mathrm{ExecF1}(y_i,y^\ast),
& \text{otherwise}.
\end{cases}
\]
Here, exactly correct executions receive an additional reward, while partially correct results are still rewarded according to ExecF1.

The final reward combines outcome quality and executability:
\[
r_i
=
\lambda r_{\mathrm{out}}(y_i)
+
(1-\lambda)r_{\mathrm{exe}}(y_i),
\]
where \(\lambda\in[0,1]\) balances outcome correctness and executability.

Through curriculum reinforcement learning, the orchestration policy is further optimized from its own rollout behaviors on challenging instances.
This completes the offline Search-to-Policy Training phase, after which the learned policy is deployed for online step-wise orchestration for \nlsql workflow.

\subsection{Online Deployment}

After offline Search-to-Policy Training, \sys is deployed as a step-wise workflow orchestration system for unseen \nlsql queries.
Given a test question \(q_{\mathrm{test}}\) and database \(\mathcal{D}_{\mathrm{test}}\), the learned policy model \(\pi_\theta\) iteratively selects actions conditioned on the current orchestration state until termination or the maximum step limit is reached.
Algorithm~\ref{alg:deployment} outlines the online procedure.

\begin{algorithm}[t]
\caption{Online Deployment of \sys}
\label{alg:deployment}
\small
\begin{algorithmic}[1]
\Require Test question \(q_{\mathrm{test}}\), test database \(\mathcal{D}_{\mathrm{test}}\), action space \(\mathcal{A}\), learned policy model \(\pi_\theta\), maximum steps \(T_{\max}\)
\Ensure Predicted \sql query \(y\)
\State \(s_0 \leftarrow \operatorname{InitState}(q_{\mathrm{test}},\mathcal{D}_{\mathrm{test}})\)
\For{\(t=1,\ldots,T_{\max}\)}
    \State \(u_t \leftarrow {\pi_\theta}(q_{\mathrm{test}},\mathcal{D}_{\mathrm{test}},\mathcal{A},s_{t-1})\) \Comment{Generate orchestration reasoning}
    \State \(e_t \leftarrow \operatorname{Extract}(u_t,\texttt{<reasoning>},\texttt{</reasoning>})\)
    \State \(a_t \leftarrow \operatorname{Extract}(u_t,\texttt{<call>},\texttt{</call>})\)
    \If{\(a_t = a_{\mathrm{term}}\)}
        \State \(y \leftarrow \operatorname{ExtractSQL}(s_{t-1})\)
        \State \Return \(y\)
    \EndIf
    \State \(o_t \leftarrow a_t(s_{t-1},\mathcal{D}_{\mathrm{test}})\) \Comment{Invoke and perform the corresponding action}
    \State \(s_t \leftarrow \operatorname{UpdateState}(s_{t-1}, e_t, a_t, o_t)\)
\EndFor
\State \(y \leftarrow \operatorname{ExtractSQL}(s_{T_{\max}})\)
\State \Return \(y\)
\end{algorithmic}
\end{algorithm}

\stitle{Step-wise Orchestration.}
At each step \(t\), \sys serializes the current orchestration state \(s_{t-1}\) into the same tagged interaction format used in training.
The policy model then generates a decision segment containing orchestration reasoning and an action call:
\(
u_t =
\texttt{<reasoning>}e_t\texttt{</reasoning>}
\texttt{<call>}a_t\texttt{</call>}.
\)

The system parses \(e_t\) from \texttt{<reasoning>} and \(a_t\) from \texttt{<call>}, and maps the parsed action name to an action module in \(\mathcal{A}\).
If \(a_t\) is a non-terminating action, the corresponding action module is invoked to produce its results and output \(o_t\), as in Eq.~\eqref{eq:traj}.
The \(o_t\) is appended to the context and incorporated into the orchestration state for the next decision.
Therefore, although the action result is not generated by the policy, it remains visible as intermediate evidence for subsequent policy decisions.

\stitle{Termination and SQL Output.}
The action space includes a termination action \(a_{\mathrm{term}}\), corresponding to \texttt{Terminate and Output}.
When the parsed \texttt{<call>} is \(a_{\mathrm{term}}\), the process terminates and extracts the current \sql candidate from the terminal orchestration state as the final prediction.
If the maximum step limit is reached, the termination is automatically triggered.
The induced workflow for the test instance is therefore:
\(
\tau_\theta=(a_1,a_2,\ldots,a_T),
\)
where \(T\le T_{\max}\) is determined online by the learned policy.
This procedure enables \sys to produce query-specific workflows with adaptive action organization, conditioned on evolving orchestration state.

\section{Experiments}\label{sec:exp}

\newcommand{\hot}{\textcolor{orange!90!red}{\tiny\faFire}\,}
\newcommand{\cold}{\textcolor{cyan!70!blue}{\tiny\faSnowflake}\,}

\begin{table*}[t]
\centering
\caption{Main comparison on BIRD-Dev. We report execution accuracy (EX, \%) on three difficulty subsets and the full set; \hot/\cold denote trained/frozen models, and \texttt{--} denotes unavailable results.}
\vspace{-1em}
\label{tab:bird_dev_results}

\begingroup
\scriptsize
\setlength{\tabcolsep}{2.2pt}
\renewcommand{\arraystretch}{1.0}
\setlength{\extrarowheight}{0pt}
\newcommand{\dynyes}{\ding{51}}
\newcommand{\dynno}{\ding{55}}

\resizebox{\textwidth}{!}{%
\begin{tblr}{
  colspec={|l|Q[c,wd=0.24\textwidth]|c|c|c|c|c|c|c|},
  cells={c,m},
  column{1}={l},
  colsep=2.2pt,
  rowsep=0pt,
  row{16-23}={bg=oursgreen},
  vline{1-Z}={abovepos=0,belowpos=0},
  hline{1,3-5,7-10,12,14,18,20,22,24}={-}{},
  hline{2,6,13,16}={1}{-}{},
  hline{2,6,13,16}={2}{-}{},
  hline{11,15,17,19,21,23}={2-3}{},
}
\textbf{Method} & \textbf{Base LLM} & \textbf{\#Params}
& \textbf{Avail.} & \textbf{Dyn. Workflow} & \textbf{Simple} & \textbf{Moderate} & \textbf{Challenge} & \textbf{Total EX} \\
CodeS~\cite{codes} & \hot StarCoder-15B & 15B & \dynyes & \dynno & 65.8 & 48.8 & 42.4 & 58.5 \\
Reasoning-SQL~\cite{reasoningsql} & \hot Qwen2.5-Coder-14B-Instruct & 14B & \dynno & \dynno & -- & -- & -- & 65.3 \\
SQL-R1~\cite{sqlr1} & \hot Qwen2.5-Coder-14B-Instruct & 14B & \dynyes & \dynno & 72.4 & 59.7 & 56.5 & 67.1 \\
OmniSQL-32B~\cite{omnisql} & \hot Qwen2.5-Coder-32B-Instruct & 32B & \dynyes & \dynno & 73.3 & 59.3 & 51.7 & 67.0 \\

CHESS~\cite{chess} & \cold Qwen2.5-Coder-32B-Instruct & 32B & \dynyes & \dynno & 73.5 & 60.9 & 53.5 & 67.8 \\
DeepEye-SQL~\cite{deepeyesql} & \cold Qwen2.5-Coder-32B-Instruct & 32B & \dynyes & \dynno & 76.5 & 63.5 & 55.6 & 70.6 \\

XiYan-32B~\cite{xiyansql} & \hot Qwen2.5-Coder-32B-Instruct & 32B & \dynyes & \dynno & 72.2 & 59.5 & 55.2 & 66.8 \\
CSC-SQL~\cite{cscsql} & \hot Qwen2.5-Coder-32B-Instruct & 32B & \dynno & \dynno & -- & -- & -- & 70.7 \\
\SetCell[r=2]{l,m} CHASE-SQL~\cite{chasesql}
& \hot Gemini-1.5-Flash & UNK
& \SetCell[r=2]{c,m}\dynno
& \SetCell[r=2]{c,m}\dynno
& \SetCell[r=2]{c,m}--
& \SetCell[r=2]{c,m}--
& \SetCell[r=2]{c,m}--
& \SetCell[r=2]{c,m}\underline{73.0} \\
& \cold Gemini-1.5-Pro & $>$200B & & & & & & \\
OpenSQL~\cite{opensql} & \hot Qwen2.5-Coder-32B & 32B & \dynno & \dynno & -- & -- & -- & 70.0 \\
Alpha-SQL~\cite{alphasql} & \cold Qwen2.5-Coder-32B-Instruct & 32B & \dynyes & \dynyes & 74.5 & \underline{64.0} & 57.2 & 69.7 \\
\SetCell[r=2]{l,m} SquRL~\cite{beyond_static_pipelines}
& \hot Qwen2.5-7B-It & 7B
& \SetCell[r=2]{c,m}\dynyes
& \SetCell[r=2]{c,m}\dynyes
& \SetCell[r=2]{c,m}--
& \SetCell[r=2]{c,m}--
& \SetCell[r=2]{c,m}--
& \SetCell[r=2]{c,m}67.6 \\
& \cold Qwen-Plus & $>$100B & & & & & & \\

\SetCell[r=2]{l,m}\textbf{\sys (Ours)} & \hot Qwen3-4B & 4B
& \SetCell[r=2]{c,m}\dynyes
& \SetCell[r=2]{c,m}\dynyes
& \SetCell[r=2]{c,m}76.5
& \SetCell[r=2]{c,m}62.2
& \SetCell[r=2]{c,m}56.3
& \SetCell[r=2]{c,m}70.3 \\
& \cold Qwen2.5-Coder-32B-Instruct & 32B & & & & & & \\
\SetCell[r=2]{l,m}\textbf{\sys (Ours)} & \hot Qwen3-8B & 8B
& \SetCell[r=2]{c,m}\dynyes
& \SetCell[r=2]{c,m}\dynyes
& \SetCell[r=2]{c,m}\underline{77.3}
& \SetCell[r=2]{c,m}63.1
& \SetCell[r=2]{c,m}\underline{57.6}
& \SetCell[r=2]{c,m}71.2 \\
& \cold Qwen2.5-Coder-32B-Instruct & 32B & & & & & & \\
\SetCell[r=2]{l,m}\textbf{\sys (Ours)} & \hot Qwen3-8B & 8B
& \SetCell[r=2]{c,m}\dynyes
& \SetCell[r=2]{c,m}\dynyes
& \SetCell[r=2]{c,m}75.8
& \SetCell[r=2]{c,m}62.6
& \SetCell[r=2]{c,m}\underline{57.6}
& \SetCell[r=2]{c,m}70.1 \\
& \cold Qwen2.5-Coder-14B-Instruct & 14B & & & & & & \\
\SetCell[r=2]{l,m}\textbf{\sys (Ours)} & \hot Qwen3-8B & 8B
& \SetCell[r=2]{c,m}\dynyes
& \SetCell[r=2]{c,m}\dynyes
& \SetCell[r=2]{c,m}\textbf{78.5}
& \SetCell[r=2]{c,m}\textbf{66.1}
& \SetCell[r=2]{c,m}\textbf{61.8}
& \SetCell[r=2]{c,m}\textbf{73.2} \\
& \cold Gemma4-31B-It & 31B & & & & & & \\
\end{tblr}
}

\endgroup
    \vspace{-1em}

\end{table*}

\subsection{Experimental Setup}
\label{sec:exp_setup}

\vspace{-.5em}
\stitle{Datasets.}
We use the following datasets for training and evaluation:
\begin{itemize}
    \item \textbf{BIRD-Train}~\cite{bird} is used to construct workflow supervision and train the orchestration policy through Search-to-Policy Learning.
    \item \textbf{BIRD-Dev}~\cite{bird} is used for the main evaluation, containing 1{,}534 challenging cross-domain questions over 11 databases.
    \item \textbf{Out-of-distribution (OOD) datasets} include Spider-Test~\cite{spider}, KaggleDBQA~\cite{kaggledbqa}, and ScienceBenchmark~\cite{sciencebenchmark}. Spider-Test is a cross-domain benchmark; KaggleDBQA contains real web databases with less standardized schemas; ScienceBenchmark focuses on complex scientific databases with highly specialized queries on policy, astrophysics, and cancer research. They are used only for OOD evaluation without dataset-specific training.
\end{itemize}

\stitle{Metrics.}
We use execution accuracy (EX) as the primary metric, which evaluates whether the predicted SQL returns the same execution result as the reference gold SQL on the database. 

\stitle{Baselines.}
We compare our \sys with the following state-of-the-art \nlsql baselines:
\begin{itemize}
    \item \textbf{Model-level training baselines} directly employ task-specific training with large-scale data to improve LLMs' \nlsql translation capabilities, including CodeS~\cite{codes}, Reasoning-SQL~\cite{reasoningsql}, SQL-R1~\cite{sqlr1}, and OmniSQL~\cite{omnisql}.
    \item \textbf{Fixed-pipeline systems} follow human-designed multi-stage procedures for \nlsql, including (i) prompting-based systems such as CHESS~\cite{chess} and DeepEye-SQL~\cite{deepeyesql}; as well as (ii) trained-model-enhanced systems: XiYan-SQL~\cite{xiyansql}, CSC-SQL~\cite{cscsql}, CHASE-SQL~\cite{chasesql}, and OpenSQL~\cite{opensql}.
    \item \textbf{Dynamic-workflow systems} try to go beyond fixed pipelines, including search-based methods such as Alpha-SQL~\cite{alphasql}, and the recent plan-then-execute orchestration method of SquRL~\cite{beyond_static_pipelines}.
\end{itemize}

\stitle{Implementation.}
All experiments are conducted on 8 Nvidia A100 GPUs. Implementations are summarized as follows:
\begin{itemize}
    \item \textbf{Model configuration.}
    \sys trains Qwen3-8B/4B policies to coordinate frozen action modules;
    \textit{unless otherwise specified, we use trained Qwen3-8B as default policy and Qwen2.5-Coder-32B-Instruct as default action LLM}.
    We also evaluate Qwen2.5-Coder-14B-Instruct and more recent Gemma4-31B-It to study the effect of action-model capacity and model family.

    \item \textbf{Policy training.}
    MCTS exploration uses \(N_{\mathrm{rollout}}=24\).
    The policy is first trained for two epochs with Stability-weighted Supervised Fine-tuning, using learning rate \(1\times10^{-5}\) with a cosine scheduler, LoRA rank \(r=64, \alpha_{\text{LoRA}}=128\).
    It is then further optimized for two epochs with Curriculum Reinforcement Learning, using learning rate \(1\times10^{-6}\), \(\lambda=0.8,\ G=4\).

    \item \textbf{Inference.}
    All LLMs are deployed with their official recommended configurations; \(T_{\max}\) is set to \(10\).
\end{itemize}

\vspace{-1em}
\subsection{Overall Performance}
\label{sec:exp_main}

\noindent\textbf{RQ1: \textit{How does \sys compare against existing state-of-the-art \nlsql systems?}}

Table~\ref{tab:bird_dev_results} reports the main comparison on BIRD-Dev.
With the 8B policy paired with Gemma4-31B-It as the action LLM, \sys achieves \textbf{73.2\%} EX, the highest result among all compared systems, outperforming the strongest baselines across model-level training, fixed-pipeline, and dynamic-workflow categories, including systems that rely on much larger proprietary models.
Notably, this result is obtained by training only the orchestration policy while keeping the action LLM frozen, rather than fine-tuning a dedicated \nlsql backbone.

To control for the same underlying models, we further compare methods built on Qwen2.5-Coder-32B-Instruct.
Under this comparable setting, \sys reaches \textbf{71.2\%} EX, outperforming fixed-pipeline systems such as DeepEye-SQL (70.6\%), CSC-SQL (70.7\%), and OpenSQL (70.0\%), the model-level training baseline OmniSQL-32B (67.0\%), and the search-based Alpha-SQL (69.7\%), all reasoning with the same backbone.
This controlled comparison thus highlights the performance advantage of \sys under the same backbone setting.

The subset results show the same trend: \sys achieves the best performance across all difficulty levels with reported subset scores.
\sys reaches \textbf{78.5\%}, \textbf{66.1\%}, and \textbf{61.8\%} on \textit{Simple}, \textit{Moderate}, and \textit{Challenging} subsets, respectively, outperforming the strongest baseline on each subset by 2.0\%, 2.1\%, and 4.6\%.
Notably, with the 8B policy, \sys obtains the same competitive performance in \textit{Challenging} set when paired with Qwen2.5-Coder-14B-Instruct and Qwen2.5-Coder-32B-Instruct (57.6\%), and reducing the policy from 8B to 4B lowers overall EX by only marginal points.
These results suggest that the learned orchestration policy can robustly coordinate action LLMs with different capacities, and that \sys's advantage comes from effective workflow orchestration rather than simply scaling model size.

\begin{figure}[t]
    \centering
    \includegraphics[width=0.99\columnwidth]{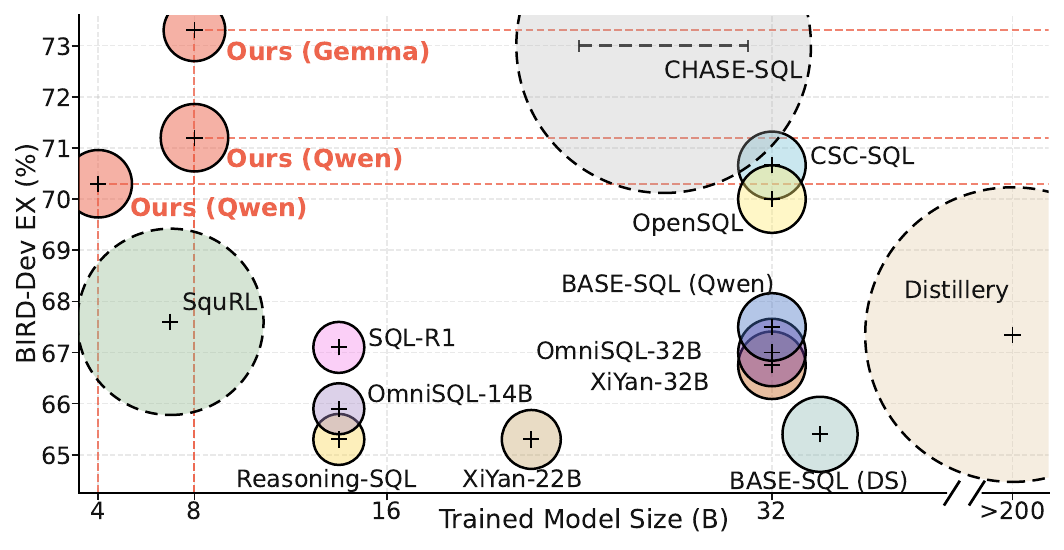}
    \vspace{-1.5em}
    \caption{BIRD-Dev EX vs. trained parameter size for training-based methods. Marker size indicates inference model size.}
    \label{fig:bubble_scatter}
    \vspace{-1.5em}
\end{figure}

Figure~\ref{fig:bubble_scatter} compares BIRD-Dev EX with the size of trained parameters across methods that adapt model weights.
\sys lies in the favorable accuracy-per-trained-parameter region: it trains only a 4B/8B orchestration policy while keeping the action LLM frozen, yet achieves superior EX.
Compared with model-level training baselines and trained-model-enhanced fixed-pipeline systems, \sys shifts learning from the \nlsql backbone to the orchestration policy, supporting the central premise that learned orchestration can be more parameter-efficient than further training a monolithic \nlsql model.

\noindent\textbf{RQ2: \textit{Does \sys learn a generalizable policy for \nlsql orchestration?}}

We then evaluate whether our trained policy to unseen database domains and query styles.
Specifically, we test \sys on Spider-Test, KaggleDBQA, and ScienceBenchmark without dataset-specific re-training, and pair the 8B policy with Qwen2.5-Coder-32B-Instruct for comparison with open-source baselines.

As shown in Table~\ref{tab:ood}, \sys achieves the best OOD average of \textbf{72.4\%}, outperforming the strongest fixed-pipeline baseline OpenSQL (71.0\%), the dynamic-workflow baseline Alpha-SQL (69.6\%), and the model-level training baseline OmniSQL-32B (69.1\%).
The gains are more pronounced on complex OOD cases involving noisy schemas or domain-specialized queries. In particular, \sys achieves the best results on KaggleDBQA and ScienceBenchmark, reaching \textbf{66.0\%} and \textbf{62.3\%}, respectively.

These results indicate that the learned orchestration policy generalizes beyond the training distribution,
suggesting that \sys does not simply fit and memorize dataset-specific patterns, but learns orchestration behavior that remains effective across diverse \nlsql settings.

\begin{table}[t]
    \centering
    \caption{OOD evaluation. Note that \sys is deployed without additional dataset-specific training.}
    \vspace{-1em}
    \label{tab:ood}
    \begingroup
    \footnotesize
    \setlength{\tabcolsep}{2.2pt}
    \renewcommand{\arraystretch}{1.0}
    \setlength{\extrarowheight}{0pt}
    \begin{adjustbox}{width=\columnwidth,center}
    \begin{tabular}{|p{0.42\columnwidth}|c|c|c|c|}
        \hline
        \textbf{Method} & \textbf{Spider.} & \textbf{Kaggle.} & \textbf{Science.} & \textbf{OOD Avg.} \\

        \hline
        \hline

        \makecell[l]{CodeS\\ + \hot StarCoder-15B} & 85.1 & 42.2 & 51.2 & 59.5 \\
        \hline
        \makecell[l]{SQL-R1\\ + \hot Qwen2.5-Coder-14B-Instruct} & 88.1 & 54.6 & 57.6 & 66.8 \\
        \hline
        \makecell[l]{OmniSQL-32B\\ + \hot Qwen2.5-Coder-32B-Instruct} & \textbf{89.8} & 56.8 & 60.6 & 69.1 \\

        \hline
        \hline

        \makecell[l]{CHESS\\ + \cold Qwen2.5-Coder-32B-Instruct} & 87.1 & 54.6 & 56.9 & 66.2 \\
        \hline
        \makecell[l]{DeepEye-SQL\\ + \cold Qwen2.5-Coder-32B-Instruct} & 88.7 & 61.6 & 60.6 & 70.3 \\
        \hline
        \makecell[l]{XiYan-32B\\ + \hot Qwen2.5-Coder-32B-Instruct} & 88.4 & 53.0 & 53.2 & 64.9 \\
        \hline
        \makecell[l]{OpenSQL\\ + \hot Qwen2.5-Coder-32B} & 88.3 & \underline{63.2} & \underline{61.5} & \underline{71.0} \\

        \hline
        \hline

        \makecell[l]{Alpha-SQL\\ + \cold Qwen2.5-Coder-32B-Instruct} & 87.8 & 62.7 & 58.2 & 69.6 \\

        \hline
        \hline

        \rowcolor{oursgreen}
        \makecell[l]{\sys\\ + \hot Qwen3-8B \\ + \cold Qwen2.5-Coder-32B-Instruct} &  \underline{88.9} & \textbf{66.0} & \textbf{62.3} & \textbf{72.4} \\
        \hline
    \end{tabular}
    \end{adjustbox}
    \endgroup
    \vspace{-1em}
\end{table}

\vspace{-0.5em}
\subsection{Efficiency and Resource Cost}
\label{sec:exp_efficiency}

\noindent\textbf{RQ3: \textit{Can \sys achieve a favorable inference-time accuracy--cost profile?}}

\begin{table}[t]
    \centering
    \caption{Per-query cost measured by action-LLM token usage under shared Qwen2.5-Coder-32B-Instruct. Parentheses denote the extra orchestration policy cost (custom Qwen3-8B).}
    \vspace{-1em}
    \label{tab:cost}
    \begingroup
    \footnotesize
    \setlength{\tabcolsep}{2.2pt}
    \renewcommand{\arraystretch}{1.0}
    \setlength{\extrarowheight}{0pt}
    \begin{adjustbox}{width=\columnwidth,center}
    \begin{tabular}{|l|c|c|c|c|c|}
        \hline
        \textbf{Method} & \textbf{EX $\uparrow$} & \textbf{Total Time (h) $\downarrow$} & \textbf{Input (k) $\downarrow$} & \textbf{Output (k) $\downarrow$} & \textbf{Avg. Cost (\$) $\downarrow$} \\
        \hline
        \hline
        CHESS & 67.8 & 21.8 & 268.0 & 37.5 & 0.109 \\
        \hline
        DeepEye-SQL & \underline{70.6} & \textbf{4.9} & \underline{20.3} & \underline{21.5} & \underline{0.024} \\
        \hline
        Alpha-SQL & 69.7 & 29.1 & 124.4 & 68.7 & 0.094 \\
        \hline
        \hline
        \rowcolor{oursgreen}
        \sys & \textbf{71.2} & \underline{5.6} & \textbf{7.1} & \textbf{12.2} & \textbf{0.012 (+0.001)} \\
        \hline
    \end{tabular}
    \end{adjustbox}
    \endgroup
    \vspace{-2em}
\end{table}

We focus on per-query workflow cost under a shared Qwen2.5-Coder-32B-Instruct action LLM backbone.
For a consistent comparison, we measure the token usage and API cost of action-LLM calls, which are available from the execution traces of representative fixed-pipeline and dynamic-workflow systems.
Costs are computed according to official Qwen API pricing.
Table~\ref{tab:cost} reports the average tokens and the corresponding cost per query.

\textit{\sys achieves the highest EX with the lowest measured per-query cost.}
As shown in Table~\ref{tab:cost}, under the shared Qwen2.5-Coder-32B-Instruct backbone, \sys achieves \textbf{71.2\%} EX with only 7.1k input tokens and 12.2k output tokens per query.
This corresponds to roughly $16\times$, $10\times$, and $2\times$ lower token usage than CHESS, Alpha-SQL, and DeepEye-SQL, respectively.
The reduced token consumption leads to an average cost of only \textbf{\$0.012} per query, with the orchestration policy adding just \text{\$0.001}, making \sys about $8\times$, $7\times$, and $2\times$ cheaper than these baselines while achieving higher EX.
In wall-clock time, \sys is substantially faster than CHESS and Alpha-SQL and remains close to DeepEye-SQL.
Overall, this shows that \sys offers a more favorable inference-time trade-off among workflow-based \nlsql methods, making efficiency an integral outcome of orchestration rather than a separate engineering optimization.

\vspace{-.0em}
\subsection{Adaptive Orchestration Analysis}
\label{sec:exp_adaptivity}

\noindent\textbf{RQ4: \textit{Does \sys improve accuracy by allocating orchestration effort according to query demands, rather than simply increasing workflow length?}}

\begin{figure}[t]
    \centering
    \includegraphics[width=0.92\columnwidth]{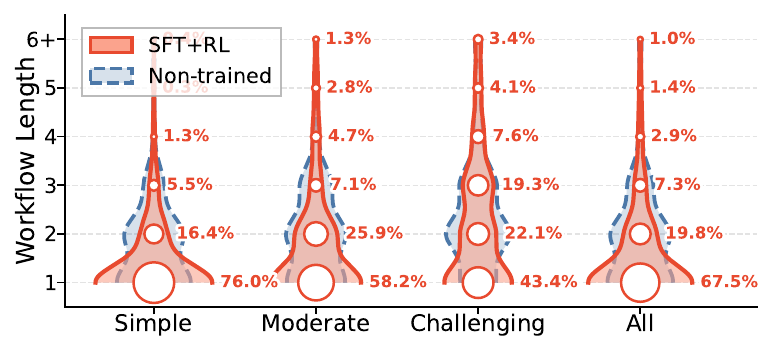}
    \vspace{-1.50em}
    \caption{Workflow-length distributions across BIRD-Dev difficulty-levels. Trained vs. Non-trained policy.}
    \label{fig:steps_violin_difficulty}
    \vspace{-1.250em}  
\end{figure}

A higher EX alone does not establish adaptivity: a policy may improve performance simply by spending more inference steps, analogous to test-time scaling.
To rule out this explanation, we compare the trained policy with the same 8B backbone before Search-to-Policy training, with both coordinating the same Qwen2.5-Coder-32B-Instruct action LLM on BIRD-Dev.
Figure~\ref{fig:steps_violin_difficulty} shows the workflow-length distributions across difficulty subsets, and Table~\ref{tab:training_ablation} reports the corresponding EX.

The trained policy improves EX on every difficulty subset, raising overall accuracy from 68.3\% to \textbf{71.2\%}, with the largest gain on \emph{Challenging} queries (51.0\% to \textbf{57.6\%}).
This improvement does not come from uniformly longer workflows: in the aggregate \emph{All} panel, the non-trained policy issues more action calls on average but obtains lower EX.
In contrast, our trained policy changes how effort is distributed across queries. Its workflow-length distribution shifts upward from Simple to Challenging queries, while the non-trained policy shows much weaker difficulty-aware adaptation.

These results suggest that Search-to-Policy learning improves not only EX but also the allocation of orchestration effort.
The trained policy achieves higher EX with fewer average action calls, while assigning additional steps more selectively to harder queries.
This behavior distinguishes adaptive orchestration from brute-force workflow expansion: the gain comes from where computation is spent, rather than simply from spending more computation overall.

\vspace{-.25em}
\subsection{Ablation Study}
\label{sec:exp_ablation}

\begin{figure}[t]
    \centering
    \includegraphics[width=0.92\columnwidth]{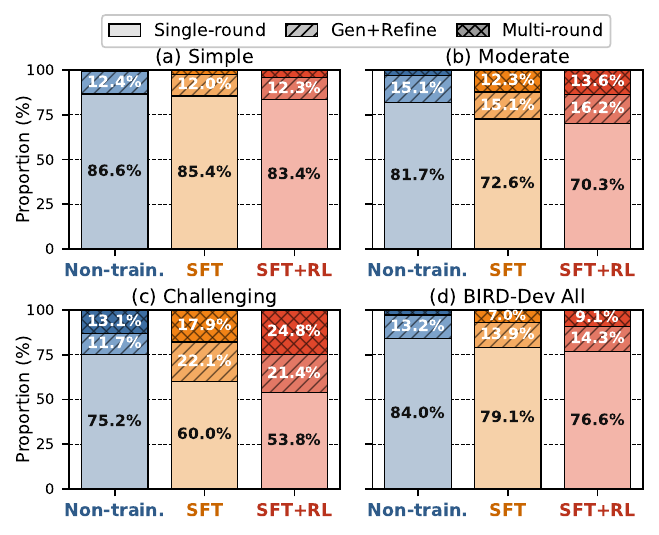}
    \vspace{-1.5em}  
    \caption{SQL-stage orchestration patterns by training stage and query difficulty in BIRD-Dev.}
    \label{fig:sql_stage_pattern}
    \vspace{-1.0em}  
\end{figure}

\begin{table}[t]
    \centering
    \caption{Ablation of SFT and RL stages on BIRD-Dev.}
    \vspace{-1.25em}
    \label{tab:training_ablation}
    \footnotesize
    \begingroup
    \setlength{\tabcolsep}{4pt}
    \renewcommand{\arraystretch}{1.08}
    \begin{tabularx}{\columnwidth}{|>{\raggedright\arraybackslash}X|c|c|c|c|}
        \hline
        \textbf{Orchestration Backbone} & \textbf{Simple} & \textbf{Moderate} & \textbf{Challenge} & \textbf{Total EX} \\
        \hline
        \hline
        \rowcolor{oursgreen} SFT + RL (full)                 & \underline{77.3} & \textbf{63.1} & \textbf{57.6} & \textbf{71.2} \\
        \hline
        SFT only, w/o RL                  & \textbf{77.5} & \underline{62.9} & 52.1 & \underline{70.7} \\
        \hline
        RL only, w/o SFT                  & 76.3 & 62.4 & \underline{55.6} & 70.2 \\
        \hline
        Non-trained policy                & 73.8 & 62.5 & 51.0 & 68.3 \\
        \hline
    \end{tabularx}
    \endgroup
    \vspace{-1.75em}
\end{table}

\noindent\textbf{RQ5: \textit{What roles do Stability-weighted SFT and Curriculum RL play in learning orchestration policies?}}

Table~\ref{tab:training_ablation} ablates these two stages while keeping the same 8B policy backbone and Qwen2.5-Coder-32B-Instruct action LLM.

\textit{Stability-weighted SFT establishes orchestration abilities, while Curriculum RL further improves difficult-query handling.}
Stability-weighted SFT raises EX from 68.3\% to 70.7\%, showing the value of curated workflow supervision.
Curriculum RL also improves over the untrained policy, suggesting that execution rewards provide useful learning signals even without supervised trajectories.
Combining both achieves the best overall EX, especially on \emph{Challenging} queries: 57.6\% versus 52.1\% for SFT-only and 55.6\% for RL-only.

Figure~\ref{fig:sql_stage_pattern} explains the gain at the behavior level.
From untrained $\to$ SFT $\to$ SFT+RL, the policy shifts from single-round SQL generation toward refinement and multi-round SQL construction, especially increasing multi-round \sql generation on \emph{Challenging} queries.
This suggests complementary roles: Stability-weighted SFT helps \sys absorb robust and diverse workflow patterns, while Curriculum RL further calibrates on challenging cases.

\vspace{1mm}
\noindent\textbf{RQ6: \textit{Are the gains of \sys attributable to policy learning, policy scale, adaptive workflow orchestration, or their combination?}}

\begin{table}[t]
    \centering
    \caption{Ablation of training, policy scale, and adaptive execution on BIRD-Dev.}
    \vspace{-1em}
    \label{tab:policy_model_ablation}
    \footnotesize
    \begingroup
    \begin{tblr}{
        width=\columnwidth,
        colspec={|X[l,m]|c|c|c|c|c|},
        cells={c,m},
        colsep=4pt,
        rowsep=1pt,
        row{2-3}={bg=oursgreen},
        vline{1-Z}={abovepos=0,belowpos=0},
        hline{1,4,6,7}={-}{},
        hline{2}={1}{-}{},
        hline{2}={2}{-}{},
        hline{3,5}={2-6}{},
    }
        \textbf{Orch. Policy} & \textbf{\#Params} & \textbf{Simple} & \textbf{Moderate} & \textbf{Challenge} & \textbf{Total EX} \\
        \SetCell[r=2]{l,m} SFT + RL & 8B & \textbf{77.3} & 63.1 & \textbf{57.6} & \textbf{71.2} \\
        & 4B & \underline{76.5} & 62.2 & \underline{56.3} & \underline{70.3} \\
        \SetCell[r=2]{l,m} Non-trained & 8B & 73.8 & 62.5 & 51.0 & 68.3 \\
        & 32B & 75.2 & \underline{63.4}  & 54.5 & 69.7 \\
        \SetCell[r=1]{l,m} Fixed workflow & -- & 74.4 & \textbf{64.2} & 52.1 & 69.2 \\
    \end{tblr}
    \endgroup
    \vspace{-2em}
\end{table}

Table~\ref{tab:policy_model_ablation} isolates the effects of policy training, policy scale, and adaptive execution, while keeping the action LLM fixed. 

\textit{Policy capacity helps but is insufficient.}
Within the same training condition, larger policies achieve higher EX: the non-trained 32B policy outperforms the non-trained 8B policy (69.7\% vs.\ 68.3\%), and the trained 8B policy outperforms the trained 4B policy (71.2\% vs.\ 70.3\%).
However, the non-trained 8B policy still underperforms the fixed workflow baseline (68.3\% vs.\ 69.2\%), indicating that policy scale alone does not guarantee reliable orchestration.

\textit{Adaptivity requires a reliable orchestration policy.}
Dynamic action selection is not automatically beneficial.
When the policy is weak, adaptivity may lead to misallocated actions and fall behind a fixed workflow.
As the policy becomes stronger or well-trained, adaptivity becomes beneficial: the non-trained 32B policy reaches 69.7\%, while the trained 4B and 8B policies further improve to 70.3\% and 71.2\%, respectively.
This suggests that adaptive orchestration should be viewed as a policy-controlled capability rather than an inherent advantage of dynamic workflows.

\textit{Policy learning is more effective than raw scaling.}
The trained 4B policy surpasses the non-trained 32B policy (70.3\% vs.\ 69.7\%), despite being roughly $8\times$ smaller.
This indicates that Search-to-Policy learning contributes orchestration behavior that cannot be explained by policy capacity alone.

Overall, the gains of \sys arise from the combination of adaptive execution and learned orchestration, with policy learning playing the central role.
Adaptive execution can underperform a fixed workflow under an unreliable policy, but becomes notably more effective when guided by a sufficiently capable learned policy.

\vspace{0.5mm}
\noindent\textbf{RQ7: \textit{How does supervision curation approach affect policy learning in Stability-weighted Supervised Fine-tuning?}}

\begin{table}[t]
    \centering
    \caption{Ablation of SFT supervision curation on BIRD-Dev. All variants use the Qwen3-8B policy backbone w/o. RL.}
    \vspace{-1.25em}
    \label{tab:sft_ablation}
    \footnotesize
    \begingroup
    \setlength{\tabcolsep}{4pt}
    \renewcommand{\arraystretch}{1.08}
    \begin{adjustbox}{width=\columnwidth,center}
    \begin{tabular}{|l|c|c|c|c|}
        \hline
        \textbf{SFT-only Policy} & \textbf{Simple} & \textbf{Moderate} & \textbf{Challenge} & \textbf{Total EX} \\
        \hline
        \hline
        \rowcolor{oursgreen}
        Stability-weighted (Top-3) & \textbf{77.5} & \underline{62.9} & \textbf{52.1} & \textbf{70.7} \\
        \hline
        Top-1 only     & \underline{77.0} & \textbf{63.3} & \underline{51.4} & \underline{70.4} \\
        \hline
        Shortest correct workflow only       & 75.4 & 59.0 & 48.6 & 67.9 \\
        \hline
        Non-trained policy                   & 73.8 & 62.5 & 51.0 & 68.3 \\
        \hline
    \end{tabular}
    \end{adjustbox}
    \endgroup
    \vspace{-1.25em}
\end{table}

To study the effect of supervision curation, Table~\ref{tab:sft_ablation} compares three strategies under the same Qwen3-8B policy backbone with RL disabled: using the explored \emph{shortest} correct workflow for each query, using the top-1 workflow with the highest stability score, and using our stability-weighted top-3.

\textit{Correct workflows are not equally effective supervision.}
Training on the \emph{shortest} correct workflow reduces total EX to 67.9\%, even below the non-trained policy (68.3\%), with larger drops on \emph{Moderate} and \emph{Challenging} queries.
This suggests that shortest-path supervision may bias the policy toward under-orchestration, as a workflow reaching the correct answer is not necessarily a robust orchestration pattern.
Using the top-1 workflow improves EX to 70.4\%, while the stability-weighted top-3 strategy achieves the best overall EX (70.7\%) and the best \emph{Challenging} performance.
These results show that SFT benefits not from imitating arbitrary correct trajectories, but from curated supervision that captures stable and diverse orchestration patterns across search rollouts.

\vspace{-0.25em}
\subsection{Case Study}
\label{sec:exp_case}

\begin{figure}[t]
    \centering
    \includegraphics[width=0.94\columnwidth]{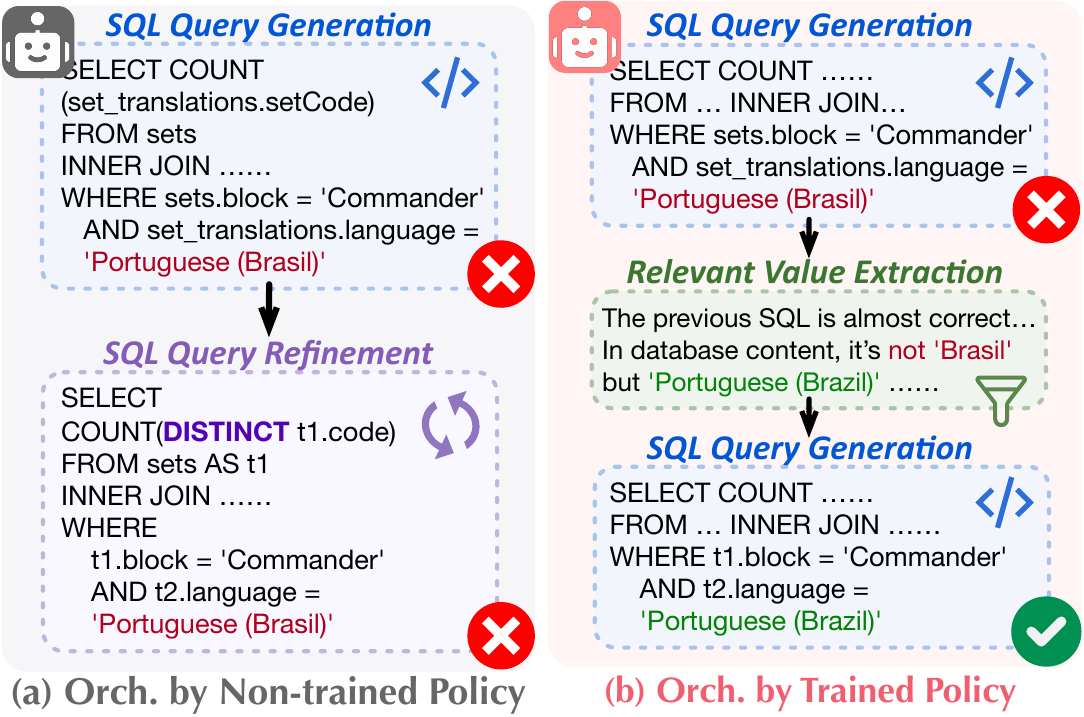}
    \vspace{-1.25em}
    \caption{Case study.}
    \label{fig:case_study}
    \vspace{-1.5em}
\end{figure}

Figure~\ref{fig:case_study} presents a representative example from BIRD-Dev (QID: 405) to demonstrate how \sys adapts its workflow.
The question requires precise value grounding: several similar surface forms exist (\eg \texttt{Brasil} and \texttt{Brazil}), but only the database-stored form can be executed correctly.
Both policies initially use the plausible but mismatched condition \texttt{Portuguese (Brasil)}, which fails to match the stored value.
After this failure, the non-trained policy naturally calls \sql refinement. However, without new value evidence, it preserves the wrong condition and leaves the error unresolved.
In contrast, the trained policy recognizes the failure as a potential value-grounding issue. It first invokes relevant-value extraction to identify \texttt{Portuguese (Brazil)}, then regenerates the query with the corrected condition.
This case shows that learned step-wise orchestration can respond to execution failure by gathering missing evidence before regeneration, rather than repeatedly refining an under-grounded query, which is consistent with the aggregate analysis in Section~\ref{sec:exp_adaptivity}.

\vspace{-0.5em}
\section{Related Work}
\vspace{-0.25em}

\stitle{LLM-based \nlsql.}
Earlier LLM-based methods mainly prompt general-purpose LLMs for \nlsql reasoning~\cite{c3sql, dinsql, supersql, tai2023exploring}.
For example, DAIL-SQL~\cite{dailsql} studies in-context learning through demonstration selection.
Some other works also focus on task-specific training, including fine-tuning~\cite{omnisql, lin2025lead}, reinforcement learning~\cite{rewardsql, reasoningsql, yao2025arctic}, and incremental pre-training~\cite{codes}, to enhance the \nlsql translation capabilities of open-source LLMs.

\stitle{Workflow Organization in \nlsql.}
Since real-world \nlsql often involves multi-stage reasoning, many systems organize the solving process as a workflow~\cite{liu2025survey, nl2sql_tutorial, dataagent_tutorial}.
CHESS~\cite{chess} and DeepEye-SQL~\cite{deepeyesql} introduce carefully designed multi-stage pipelines for schema linking, \sql generation, revision, and selection to improve \nlsql performance with prompting-based LLMs~\cite{nl2sqlbugs, li2026dpc}.
Some systems further integrate trained reasoning models into such pipelines~\cite{xiyansql, chasesql, sheng2025base} or use specially trained LLMs for every individual stage~\cite{opensql, cscsql, qinroute, he2025star, agentarscalesql}.
However, these pipelines usually follow predefined stage orders, limiting their adaptivity to different query demands and intermediate feedback.
Some methods thus explore dynamic workflow organization for \nlsql, including complexity-based routing~\cite{elliesql} and inference-time MCTS search over reasoning paths~\cite{alphasql, liao2026learnat, mctssql, lyu2025sql}. More recently, SquRL~\cite{beyond_static_pipelines} attempts to plan a complete query-specific workflow before execution.
While these methods improve system flexibility, they either incur substantial online search costs or still fix the workflow once execution begins.
In contrast, \sys introduces Search-to-Policy Learning to train a step-wise orchestration policy, enabling adaptive action selection and workflow revision during inference without extensive online search.

\vspace{-0.5em}
\section{Conclusion}

We propose \sys, a step-wise orchestration learning framework for \nlsql that adapts \nlsql workflow composition.
\sys introduces Search-to-Policy Learning, which combines MCTS-based workflow exploration and curation, Stability-weighted Supervised Fine-tuning, and Curriculum Reinforcement Learning to transform broad offline search into a deployable orchestration policy.
Extensive experiments demonstrate that \sys achieves strong performance and generalization, with adaptive \nlsql orchestration.

\bibliographystyle{ACM-Reference-Format}
\bibliography{references}

\end{document}